\documentclass[12pt]{article}

\usepackage{amssymb}
\usepackage{axodraw}

\def\be{\begin{equation}}
\def\ee{\end{equation}}
\def\bea{\begin{eqnarray}}
\def\eea{\end{eqnarray}}
\def\cN{{\cal N}}
\def\dfrac#1#2{{\displaystyle\frac{#1}{#2}}}
\def\tfrac#1#2{{\textstyle\frac{#1}{#2}}}
\def\dint{{\displaystyle\int}}

\def\Tr{{\rm Tr}}
\def\nn{\nonumber \\}

\newcommand{\p}[1]{(\ref{#1})}
\def\theequation{\arabic{section}.\arabic{equation}}

\makeatletter
\long\def\@makecaption#1#2{{%
  \small
  \vskip\abovecaptionskip
  \sbox\@tempboxa{#1. #2}%
  \ifdim \wd\@tempboxa >\hsize
    #1. #2\par
  \else
    \global \@minipagefalse
    \hb@xt@\hsize{\hfil\box\@tempboxa\hfil}%
  \fi
  \vskip\belowcaptionskip}}
\makeatother

\begin{document}
\begin{titlepage}
\begin{flushright}
ITP-UH-23/05\\
\tt hep-th/0511234
\end{flushright}
\vspace{5mm}
\begin{center}
{\Large\bf
Renormalizability of non-anticommutative  \\
\vspace{0.2cm}
$\cN=(1,1)$ theories with singlet deformation}
 \\[1cm]
 {\bf
 I.L. Buchbinder $^+$\footnote{joseph@tspu.edu.ru},
 E.A. Ivanov $^\dag$\footnote{eivanov@theor.jinr.ru},
 O. Lechtenfeld $^\ddag$\footnote{lechtenf@itp.uni-hannover.de},\\[2mm]
 I.B. Samsonov $^{*}$\footnote{samsonov@phys.tsu.ru},
 B.M. Zupnik $^\dag$\footnote{zupnik@theor.jinr.ru}}\\[3mm]
 {\it $^+$ Department of Theoretical Physics, Tomsk State Pedagogical
 University,\\ Tomsk 634041, Russia\\[2mm]
 $^\dag$ Bogoliubov Laboratory of Theoretical Physics, Joint Institute for
 Nuclear Research, Dubna, 141980 Moscow Region, Russia\\[2mm]
 $^\ddag$ Institut f\" ur Theoretische Physik, Universit\" at Hannover,\\
 Appelstra\ss e 2, D-30167 Hannover, Germany\\[2mm]
 $^*$ Laboratory of Mathematical Physics, Tomsk Polytechnic
 University,\\ 30 Lenin Ave, Tomsk 634050, Russia
 }
 \\[0.8cm]
\bf Abstract
\end{center}
We study the quantum properties of two theories with a
non-anticommutative (or nilpotent) chiral singlet deformation of
$\cN=(1,1)$ supersymmetry: the abelian model of a vector gauge
multiplet and the model of a gauge multiplet interacting with a
neutral hypermultiplet. In spite of the presence of a
negative-mass-dimension coupling constant (deformation parameter),
both theories are shown to be finite in the sense that the full
effective action is one-loop exact and contains finitely many
divergent terms, which vanish on-shell. The $\beta$-function for
the coupling constant is equal to zero. The divergencies can all
be removed off shell by a redefinition of one of the two scalar
fields of the gauge multiplet. These notable quantum properties
are tightly related to the existence of a Seiberg-Witten-type
transformation in both models.

\end{titlepage}
\setcounter{footnote}{0}

\section{Introduction and summary}
Recently, there was a string-theory motivated \cite{Strings} surge
of interest in non-anticommutative deformations of Euclidean
superspaces and the corresponding deformed supersymmetric field
theories \cite{Seiberg} -- \cite{B}. The non-anticommutative (or
``nilpotently deformed'') field theories, introduced and studied in
\cite{Strings,Seiberg}, provide an effective description of the
low-energy dynamics of D-branes in superstring theory in the
presence of constant graviphoton flux. Their possible physical
applications, as established in \cite{Strings}, are related, e.g.,
to modifications of the glueball superpotential and the expectation
value for the glueball field in $\cN=1$ supersymmetric field
theories. The non-anticommutativity is also of interest from the
phenomenological point of view, since it provides a new mechanism
for the explicit partial breaking of supersymmetry. When such a
deformation is turned on, only a part (generically, half) of the
original supersymmetry remains realized in a standard way and
can still be regarded as a symmetry of the deformed
field-theoretical model. The study of field theories in
non-anticommutative $\cN = (1/2,1/2)$ superspace, initiated by ref.
\cite{Seiberg}, was soon extended to the case of non-anticommutative
$\cN = (1,1)$ superspace \cite{ILZ,Ferrara}.

Non-anticommutative field theories require new geometrical structures for
their concise formulation, which is one more source of interest
in such theories. The non-an\-ti\-com\-mu\-ta\-ti\-vi\-ty brings
additional parameters which lead to the deformation of the anticommutation
relations for the Grassmann
coordinates of superspace, e.g. $\{\theta^\alpha,\theta^\beta
\}_\star=C^{\alpha\beta}$ in $\cN =(1/2,1/2)$ superspace \cite{Seiberg} or
$\{\theta^\alpha_i,\theta^\beta_j \}_\star=C^{\alpha\beta}_{ij}$
for $\cN = (1,1)$ superspace \cite{ILZ}, with $C^{\alpha\beta}$ and
$C^{\alpha\beta}_{ij}$ being constant matrices.
The deformed models are described in superspace by introducing the
$\star$-multiplication of superfields which is associative but
non(anti)commutative \cite{Superstar}.
This means that the deformed models are
obtained from the conventional supersymmetric ones just by
replacing the usual multiplication of superfields by the
$\star$-product. In such a way, the Wess-Zumino and super
Yang-Mills (SYM) models in non-anticommutative $\cN=(1/2,1/2)$
Euclidean superspace were constructed in \cite{Seiberg}. The
non-anticommutative versions of the $\cN=(1,1)$ vector
multiplet and hypermultiplet in the $\cN = (1,1)$ harmonic superspace approach
were pioneered in \cite{ILZ,Ferrara}. A more detailed treatment of these
$\cN=(1,1)$ models, including the analysis of the component structure of the
actions respecting unbroken $\cN =(1,0)$ supersymmetry, was undertaken in
\cite{FILSZ} -- \cite{CILQ}.

The non-anticommutative field theories reveal surprising quantum properties.
An amazing feature is that, for all currently studied examples,
the renormalizability properties of deformed theories are not spoiled as
compared to the case without deformation.
More specifically, non-anticommutative $\cN = (1/2,0)$ supersymmetric
Wess-Zumino and Yang-Mills models
were proved to be renormalizable \cite{Britto} -- \cite{Grisaru2}.
From the field-theoretical point
of view this property looks rather mysterious since such models contain
the parameters of non-anticommutativity $C^{\alpha\beta}$
with mass dimension $-1$ and, by naive power-counting arguments,
should be divergent at any order of perturbation theory. However, a key feature of
non-anticommutativity is that all such models are consistent only
in Euclidean space where the reality properties radically differ
from those in Minkowski space. An important manifestation of this difference
in the quantum computations is that the new
vertices appearing with the parameters $C^{\alpha\beta}$ are not
accompanied by their conjugates and, for this reason, only finitely many
divergent Feynman
graphs with new divergencies appear. For example, for
the non-anticommutative Wess-Zumino model it was shown that only a single new
divergent term should be added to the classical action, and
the model where such an extra term is added from the very beginning
is renormalizable in the usual sense \cite{Britto,0306237,Romagnoni,Ber}. The
non-anticommutative super Yang-Mills model was also proved to be renormalizable
using component field formulations \cite{Ber,Lunin,ir,Jack} and superspace techniques
\cite{Grisaru2}.
Note that, in $\cN=(1/2,0)$ gauge theories, quantum computations in components
also produce new field structures which are not present in the classical action
but, in contrast to the Wess-Zumino model, can be
removed from the effective action by a simple
shift of the gaugino field \cite{Jack}. Thus, $\cN=(1/2,0)$ non-anticommutative
theories provide interesting examples of renormalizable field theories
with dimensionful coupling constants, i.e. the deformation parameters.

The study of theories with non-anticommutative deformations of $\cN=(1,1)$
supersymmetry is more involved. In particular, there are different types
of chiral deformation of $\cN=(1,1)$ superspace related to different
choices of constants $C^{\alpha\beta}_{ij}=\{\theta^\alpha_i,\theta^\beta_j \}_\star\,$.
The simplest case $C^{\alpha\beta}_{ij}=2 I\varepsilon^{\alpha\beta}
\varepsilon_{ij}$ is called the singlet deformation \cite{ILZ,Ferrara}. The
$\cN=(1,1)$ classical models with singlet non-anticommutative deformation have been
constructed in \cite{FILSZ,ILZ2,Araki}. Note that the singlet
non-anticommutative deformations of $\cN=(1,1)$ superspace also
emerge from string theory considered on the axionic background
\cite{FILSZ}. The case of
non-singlet deformations was considered in \cite{Araki1,CILQ}, where the
structure of the classical action of the super Yang-Mills model was
studied in some detail. In this paper we will analyze
only models with singlet deformation, which can be described
using the following $\cN=(1,1)$ superfield $\star$-product \cite{ILZ},
\be
A\star B= A\exp\left(-I\varepsilon^{\alpha\beta}\varepsilon_{ij}
\overleftarrow{Q}_\alpha^i \overrightarrow{Q}_\beta^j\right)B,
\label{in1}
\ee
where $Q^i_\alpha$ are the $\cN=(1,1)$ supercharges and $I$ is the parameter
of singlet non-an\-ti\-com\-mu\-ta\-ti\-vi\-ty. The multiplication (\ref{in1}) breaks
$\cN=(1,1)$ supersymmetry by half, i.e. down to $\cN=(1,0)$. Nevertheless,
this deformation preserves chirality and Grassmann harmonic
analyticity \cite{ILZ} and therefore suits well for the use of $\cN=2$ harmonic
superspace techniques. The classical superfield actions for non-anticommutative
$\cN=(1,1)$ models of the vector multiplet and the hypermultiplet were constructed
in harmonic superspace in \cite{ILZ}. The component structure of these
actions was studied extensively \cite{FILSZ,ILZ2,Araki}.
However, the quantum aspects of such models have never been studied. In particular,
the problem of
the renormalizability of these models and the problem of
constructing the effective
actions have not been addressed so far. The study of these issues is of
substantial interest
since the renormalizability of deformed $\cN=(1,1)$
models is not evident and  we can expect that the effective action in the case
under consideration will possess a number of very nontrivial properties.

By this paper we begin a systematic study of quantum aspects of the
non-an\-ti\-com\-mu\-ta\-ti\-ve models with deformed $\cN=(1,1)$
supersymmetry. Here we answer affirmatively the question of renormalizability for
non-an\-ti\-com\-mu\-ta\-ti\-ve models of the abelian $\cN=(1,1)$ vector multiplet
\footnote{Although the model under consideration has U(1) gauge symmetry,
it is interacting due to non-an\-ti\-com\-mu\-ta\-ti\-vi\-ty.},
with and without adding a neutral hypermultiplet, in the case of a
singlet deformation of the form (\ref{in1}).
The renormalizability of these models can be formally established by
applying the Wilsonian renormalization group arguments, developed
previously for $\cN=(1/2,0)$ supersymmetric models in \cite{Ber}.
To find the structure of divergent terms in the effective action
we perform the one-loop quantum computations and observe several specific
properties of the two models. First, the divergent terms appear only
in the vector loop in the SYM model or in the scalar loop in the
hypermultiplet model. The external lines must carry only
the scalar $\bar\phi$ belonging to the vector multiplet. The
fermionic contributions are trivial. Second, all divergent
contributions combine in two gauge invariant
and $\cN=(1,0)$ supersymmetric expressions depending on
the single scalar field $\bar\phi$ only and vanishing at the classical
equations of motion. These two divergent terms
represent new interactions which were not present in the
classical actions of the considered models. However, they do not spoil
the renormalizability
since they can be completely absorbed into a redefinition of the
second scalar field $\phi$. This situation is very similar to the
$\cN=(1/2,0)$ gauge model studied in \cite{Jack}, where it was shown that the
non-anticommutative interaction also generates new terms in the effective action
which can be eliminated from the
theory by a shift of the gaugino field. Third, we demonstrate that
the appropriate change of fields in the classical actions
(a Seiberg-Witten-like map)\cite{FILSZ,ILZ2} allows one to completely avoid
any divergence in the effective action. This fact emphasizes that the
divergencies are unphysical from the standard point of view.

%We would like to point out that all these specific quantum features of
%the considered  models were revealed by direct quantum computations
%based on the Feynman diagram method. Thus the latter provides us with a detailed information
%on the quantum structure of these models, which is more than just a proof of their
%renormalizability.
%
%As a result,
%the considered $\cN=(1,0)$ non-anticommutative models prove
%to be not only renormalizable, but actually finite, since their parameters
%do not require any infinite renormalization.
%Note that these are direct quantum computations

Despite the disadvantage that the considered theories have to be regarded
as a sort of toy models, they are relevant for the quantum structure of
non-abelian non-an\-ti\-com\-mu\-ta\-ti\-ve SYM theory, since they appear as a U(1)
sector in non-anticommutative U($N$) gauge theories. Hence, the proof of
renormalizability of the abelian vector multiplet and hypermultiplet
models is a prerequisite for the analogous study of the deformed U($N$)
$\cN=(1,1)$ gauge theories. Moreover, the results obtained support
the hypotheses that the non-anticommutative deformations of
supersymmetry cannot spoil the renormalizability of supersymmetric
theories and provide a way of construction of renormalizable
supersymmetric models with partially broken supersymmetry.

The paper is organized as follows. In Section 2 we
compute the one-loop effective action in the
non-anticommutative $\cN=(1,1)$ abelian gauge theory
formulated in terms of component fields and prove the
renormalizability of this model. In Section 3 we prove, in the
same way as for the pure gauge theory, the renormalizability of
the non-anticommutative deformation of a coupled system
of neutral hypermultiplet and abelian gauge multiplet.
In Section 4 we study the interplay between the Seiberg-Witten map and
the problem of renormalizability of the considered models.
Section 5 contains the conclusions and some discussion of the results
obtained. In the Appendix we collect the formulas for the regularization of
divergent momentum integrals which are met in the calculations of the
effective action, and we also list the corresponding Feynman
diagrams.

We follow the conventions and notation of refs. \cite{FILSZ,ILZ2}.

\setcounter{equation}0
\section{Renormalizability of non-anticommutative abelian $\cN = (1,0)$
supergauge theory}
\label{sec1}
The classical superfield action of $\cN=(1,1)$ non-anticommutative U(1) gauge model was
given in \cite{ILZ,Ferrara,FILSZ}
\be
S_{SYM}=\frac14\int d^8z_c du\, {\cal W}^2\,.
\label{s1.1}
\ee
Here $\cal W$ is the $\cN=(1,1)$ superfield strength which is covariantly chiral,
$\bar{\nabla}^i_{\dot\alpha}\star{\cal W}$ $=$ \break
$(\bar D^i_{\dot\alpha}+\bar A^i_{\dot\alpha}\star)
{\cal W}=0\,$,
$d^8z_c=d^4x_c d^4\theta$ is the integration measure of the chiral
superspace and $du$ stands for the integration over the harmonic variables.
The action (\ref{s1.1}) is invariant under the deformed U(1) gauge transformations
\be
\delta{\cal W}=[{\cal W},\Lambda]_\star={\cal
W}\star\Lambda-\Lambda\star{\cal W},
\label{s1.2}
\ee
where $\Lambda$ is an arbitrary analytic superfield and the $\star$-product
was defined in \p{in1}.

The covariantly chiral superfield $\cal W$ can be decomposed into
usual chiral superfields
\be
{\cal W}={\cal A}+\bar\theta^+_{\dot\alpha}\tau^{-\dot\alpha}
+(\bar\theta^+)^2\tau^{-2}\,,
\label{s1.3}
\ee
with $\bar D^i_{\dot\alpha}{\cal A} =\bar D^i_{\dot\alpha}\tau^{-\dot\alpha}
=\bar D^i_{\dot\alpha}\tau^{-2}=0\,$.
It is easy to demonstrate that only the superfield $\cal A$ contributes
to the action (\ref{s1.1}) \cite{FILSZ}
\be
S_{SYM}=\frac14\int d^8z_c du\, {\cal A}^2.
\label{s1.4}
\ee

The component structure of $\cal A$ was found in \cite{FILSZ}
\begin{eqnarray}
{\cal A}(z_c,u)&=&\displaystyle[\phi+\frac{4IA_mA_m}{1+4I\bar\phi}
+\frac{16I^3(\partial_m\bar\phi)^2}{1+4I\bar\phi}]
\nonumber\\&&
+2\theta^+[\Psi^-+\frac{4I(\sigma_m\bar\Psi^-)A_m}{1+4I\bar\phi}]
-\frac{2\theta^-}{1+4I\bar\phi}[\Psi^++\frac{4I(\sigma_m\bar\Psi^+)A_m}{1+4I\bar\phi}]
\nonumber\\&&
+(\theta^+)^2[\frac{8I(\bar\Psi^-)^2}{1+4I\bar\phi}+{\cal D}^{--}]
+\frac{(\theta^-)^2}{(1+4I\bar\phi)^2}[\frac{8I(\bar\Psi^+)^2}{1+4I\bar\phi}+{\cal
D}^{++}]
\nonumber\\&&
-\frac{2(\theta^+\theta^-)}{1+4I\bar\phi}[\frac{8I(\bar\Psi^+\bar\Psi^-)}{1+4I\bar\phi}
+{\cal D}^{+-}]
+(\theta^+\sigma_{mn}\theta^-)(F_{mn}-\frac{8I\partial_{[m}\bar\phi A_{n]}}{
1+4I\bar\phi})\nonumber\\&&\displaystyle
+2i(\theta^-)^2\theta^+\sigma_m\partial_m\frac{\bar\Psi^+}{1+4I\bar\phi}
+2i(\theta^+)^2(1+4I\bar\phi)\theta^-\sigma_m\partial_m\frac{\bar\Psi^-}{1+4I\bar\phi}
\nonumber\\&&\displaystyle
-(\theta^+)^2(\theta^-)^2\square\bar\phi\,.
\label{s1.5}
\end{eqnarray}
Here ${\cal D}^{++}={\cal D}^{kl}u_k^+ u_l^+$, ${\cal D}^{+-}=
{\cal D}^{kl}u_k^+ u_l^-\,$, $\phi,\ \bar\phi$ are the scalar fields, $\Psi^\pm_\alpha=\Psi^i_\alpha
u^\pm_i$, $\bar\Psi^\pm_{\dot\alpha}=\bar \Psi^i_{\dot\alpha}
u^\pm_i$ are the spinors, $A_m$ is the vector field, ${\cal D}^{kl}$ is the auxiliary field.
Substituting the expression (\ref{s1.5}) into the action
(\ref{s1.4}) we find
\begin{eqnarray}
S_{SYM}&=&S_\phi+S_{\Psi}+S_A,\label{s1.6}\\
S_\phi&=&-\frac12\int d^4x\square\bar\phi\left[\phi+\frac{4IA_mA_m}{1+4I\bar\phi}
+\frac{16I^3\partial_m\bar\phi\partial_m\bar\phi}{1+4I\bar\phi}\right],\label{s1.7}\\
S_\Psi&=&i\int d^4x\left(
\Psi^{i\alpha}+\frac{4IA_m\sigma_m{}^{\alpha}{}_{\dot\alpha}\bar\Psi^{i\dot\alpha}}{
1+4I\bar\phi}
\right)(\sigma_n)_{\alpha\dot\beta}\partial_n
\left(\frac{\bar\Psi_i^{\dot\beta}}{1+4I\bar\phi} \right)
\nonumber\\
&&+\frac14\int d^4x
\frac1{(1+4I\bar\phi)^2}
\left(\frac{8I\bar\Psi^i_{\dot\alpha}\bar\Psi^{j\dot\alpha}}{1+4I\bar\phi}
+{\cal D}^{ij}\right)\left(\frac{8I\bar\Psi_{i\dot\alpha}\bar\Psi_j^{\dot\alpha}}{1+4I\bar\phi}
+{\cal D}_{ij}\right),\label{s1.8}\\
S_A&=&\int d^4x\left[-\frac12 A_n\square A_n
-\frac12\partial_m A_m \partial_n A_n
+\frac12 A_nA_n\square\ln(1+4I\bar\phi)\right.\nonumber\\&&
-\varepsilon_{mnrs}\partial_rA_sA_n\partial_m\ln(1+4I\bar\phi)
+\frac12
A_nA_n\partial_m\ln(1+4I\bar\phi)\partial_m\ln(1+4I\bar\phi)\nonumber\\&&\left.
-\frac12 A_mA_n\partial_m\ln(1+4I\bar\phi)\partial_n\ln(1+4I\bar\phi)
+\partial_n A_m A_n\partial_m\ln(1+4I\bar\phi)\right].
\label{s1.9}
\end{eqnarray}
At this step it is convenient to eliminate the auxiliary field
${\cal D}^{ij}$ using its classical equation of motion
${\cal D}^{ij}=-8I\bar\Psi_{\dot\alpha}^i
\bar\Psi^{j\dot\alpha}/(1+4I\bar\phi)$, then the terms
in the second line of (\ref{s1.8}) disappear and the action $S_\Psi$
simplifies to
\be
S_\Psi=i\int d^4x\left(
\Psi^{i\alpha}+\frac{4IA_m\sigma_m{}^{\alpha}{}_{\dot\alpha}\bar\Psi^{i\dot\alpha}}{
1+4I\bar\phi}
\right)(\sigma_n)_{\alpha\dot\beta}\partial_n
\left(\frac{\bar\Psi_i^{\dot\beta}}{1+4I\bar\phi} \right).
\label{s1.10}
\ee
Note that model (\ref{s1.6}) is formulated in the Euclidean rather
then Minkowski space.
This means that the fields $\phi$, $\bar\phi$ and $\Psi^i_\alpha$,
$\bar\Psi^i_{\dot\alpha}$ are not conjugated to each other.

Before continuing with the quantization of the model (\ref{s1.6}) we
would like to adduce here some intuitive arguments in favour of the
renormalizability of this theory. Following the work \cite{Ber}, we
will use the Wilsonian approach \cite{Wilson} which ensures
the renormalizability of lagrangians containing a finite number of
local operators with scaling dimensions not greater than four. The
lagrangian (\ref{s1.6}) contains a finite number of local operators
(interaction terms), but the dimensions of some of these operators
are greater than four. For example, the interaction terms
\be
\frac{\square\bar\phi A_mA_m}{1+4I\bar\phi},\qquad
\frac{\square\bar\phi\partial_m\bar\phi\partial_m\bar\phi}{1+4I\bar\phi}
\label{s1.10_1}
\ee
in (\ref{s1.7}) have mass dimensions 5 and 7,
respectively. Therefore, the arguments of
Wilsonian approach cannot be directly applied to the action (\ref{s1.6}).
However, in \cite{Ber} it was proposed to ascribe
non-standard scaling dimensions to the fields involved. Following this way,
let us suppose that the dimensions of the coordinates of $\cN=2$ superspace are
\be
[\theta^\alpha_i]=0,\qquad [\bar\theta^{\dot\alpha i}]=-1,\qquad [x^m]=-1.
\label{s1.10_}
\ee
Then it is easy to see that the dimensions of physical component fields
should be as follows,
\be
\begin{array}{lll}
{}[\phi]=2,\quad &[\bar\phi]=0,\quad& [A_m]=1,\\
{}[\Psi^{i\alpha}]=0,& [\bar\Psi^{i\dot\alpha}]=1
\end{array}
\label{s1.10_2}
\ee
and the parameter of non-anticommutativity is
dimensionless, $[I]=0$. When the asymmetrical scaling dimensions
(\ref{s1.10_2}) are assumed, the operators (\ref{s1.10_1}), as
well as other interaction terms in (\ref{s1.6}), acquire the
desired mass dimension 4. Therefore, such a model has to be
renormalizable on formal grounds by applying the Wilsonian
argument. Note that one can choose the asymmetrical scaling
dimensions (\ref{s1.10_2}) just because the fields $\phi$ and
$\bar\phi$ as well as $\Psi^{i\alpha}$ and
$\bar\Psi^{i\dot\alpha}$ are not related to each other by
conjugation. The above argument based
on the relations (\ref{s1.10_2}) is useful only for establishing
the very fact of renormalizability and does not provide any
specific rules for performing the renormalization and/or
clarifying its details. %We will show that the model under
%consideration enjoys a quite special renormalization structure.
Having established the renormalizability of the model in
principle, we henceforth adopt the standard canonical dimensions
for all fields.

Now we are going to consider the quantization of the model (\ref{s1.6})
in order to study the structure of divergent terms and the details of
renormalization procedure. The action (\ref{s1.6}) is invariant under the
following residual gauge transformations
\be
\begin{array}{ll}
\delta\phi=-8IA_m\partial_m \lambda,  &\delta\bar\phi=0,\\
\delta\Psi^k_\alpha=-4I(\sigma_m\bar\Psi^k)_\alpha\partial_m\lambda,\quad
& \delta \bar\Psi^k_{\dot\alpha}=0,\\
\delta A_m=(1+4I\bar\phi)\partial_m\lambda,&
\delta {\cal D}^{kl}=0\,,
\end{array}
\label{s1.11}
\ee
with $\lambda$ being the gauge parameter.
Note that the gauge field $A_m$ has a non-standard transformation law. However,
after the redefinition $A_m\rightarrow a_m=A_m/(1+4I\bar\phi)$
the new field $a_m$ has the standard gauge transformation law $\delta
a_m=\partial_m\lambda$. Therefore, the standard Lorentz gauge
fixing condition reads $\partial_m a_m=0$, or
\be
\partial_m\frac{A_m}{1+4I\bar\phi}=0.
\label{s1.12}
\ee

Further we follow the routine Faddeev-Popov procedure to fix the
gauge freedom in the functional integral.
Let us introduce the corresponding gauge-fixing function
\be
\chi=\partial_m\frac{A_m}{1+4I\bar\phi}=
\frac{\partial_m A_m-A_m G_m}{1+4I\bar\phi},
\label{s1.13}
\ee
where
\be
G_m(x)=\partial_m \,\ln[1+4I\bar\phi(x)]\,.
\label{s1.14}
\ee
The function (\ref{s1.13}) transforms under the gauge
transformations (\ref{s1.11}) as
\be
\delta\chi=\partial_m\frac{\delta
A_m}{1+4I\bar\phi}=\square\lambda.
\label{s1.15}
\ee
Therefore the action for the ghost fields is just the action of
free scalars
\be
S_{FP}=\int d^4x\, b\square c\,.
\label{s1.16}
\ee
The generating functional for the Green's functions is now defined
as
\footnote{Note that the functional integral in the Euclidean space
is defined as $\int {\cal D}\Phi e^{-\frac12 S[\Phi]}$ as compared with
the Minkowski space definition $\int {\cal D}\Phi e^{\frac i2 S[\Phi]}$.}
\be
Z[J]=\int{\cal D}(\phi,\bar\phi,\Psi,\bar\Psi,A_m,b,c)
\delta(\chi-\tfrac{\partial_m A_m-A_m G_m}{1+4I\bar\phi})
e^{-\frac12( S_{SYM}+S_{FP}+S_J)}\,,
\label{s1.17}
\ee
where
\be
S_J=\int d^4x[\phi J_\phi+\bar\phi J_{\bar\phi}
+\Psi^i_\alpha (J_{\Psi})_i^\alpha+
\bar\Psi_{i\dot\alpha} (J_{\bar\Psi})^{i\dot\alpha}+ A_m (J_A)_m]
\label{s1.18}
\ee
and $J_\phi$, $J_{\bar\phi}$,
$(J_\Psi)_i^\alpha$, $(J_{\bar\Psi})^{i\dot\alpha}$, $(J_A)_m$ being sources
associated with the fields
$\phi$, $\bar\phi$, $\Psi^i_\alpha$, $\bar\Psi_{i\dot\alpha}$,
$A_m$. We have inserted into (\ref{s1.17})
the functional delta-function which fixes the gauge degrees of freedom in
the functional integral over the gauge fields. This delta-function
can be easily written in the gaussian form by
averaging (\ref{s1.17}) with the factor
\be
1=\int {\cal D}\chi e^{-\frac\alpha2\int d^4x  \chi^2(1+4I\bar\phi)^2}
={\rm Det}^{-1/2}[\delta^4(x-x')(1+4I\bar\phi)^2]\,.
\label{s1.19}
\ee
The functional integral (\ref{s1.19}) generates the following
gauge-fixing action
\bea
S_{gf} &=&\frac\alpha2\int d^4x(\partial_m A_m-A_m G_m)^2 \nn
&=&\frac\alpha2\int d^4x[(\partial_m A_m)^2-2\partial_m A_m A_n G_n
+A_mA_n G_mG_n]\,.
\label{s1.20}
\eea
Here $\alpha$ is an arbitrary parameter. For simplicity, in sequel we set
$\alpha=1\,$. As a result, the generating functional
(\ref{s1.17}) can be represented in the following form
\be
Z[J]=\int{\cal D}(\phi,\bar\phi,\Psi,\bar\Psi,A_m,b,c)
e^{-\frac12(S_{\rm tot}+S_{FP}+S_J)}\,,
\label{s1.21}
\ee
where
\begin{eqnarray}
S_{\rm tot}&=&S_{SYM}+S_{gf}\nonumber\\
&=&-\frac12\int d^4x\square\bar\phi(\phi
+4I^2\partial_m\bar\phi G_m)\nonumber\\
&&+\,i\int d^4x\left(
\Psi^{i\alpha}+\frac{4IA_m\sigma_m{}^{\alpha}{}_{\dot\alpha}\bar\Psi^{i\dot\alpha}}{
1+4I\bar\phi}
\right)(\sigma_n)_{\alpha\dot\beta}\partial_n
\left(\frac{\bar\Psi_i^{\dot\beta}}{1+4I\bar\phi} \right)\nonumber\\
&&
-\int d^4x\left[\frac12 A_n\square A_n- A_n G_m\partial_n A_m+
A_n G_n\partial_m A_m+\varepsilon_{mnrs}G_mA_n\partial_r
A_s\right].
\label{s1.22}
\end{eqnarray}

The functional integral (\ref{s1.21}) with the action (\ref{s1.22})
requires several comments.

\begin{enumerate}
\item The ghost fields $b,\ c$ enter the action only through
their kinetic term. Hence, they fully decouple and can be integrated
out.

\item The action (\ref{s1.22}) defines the propagators and
vertices, all what we need for performing quantum computations
in the model. Upon a careful examination of the Feynman rules,
one can prove the following statements (see Appendix A.2):
\begin{itemize}
\item[{\bf i}.] The one-loop effective action in the model is exact since it
is impossible to construct any higher loop diagrams;
\item[{\bf ii}.] The fermionic fields $\Psi^i_\alpha,\
\bar\Psi^i_{\dot\alpha}$ do not produce any quantum corrections to
the effective action (excepting for tree diagrams);
\item[{\bf iii}.] Only the field $\bar\phi$ can appear at the external
legs;
\item[{\bf iv}.] The only contribution to the effective action comes
from the vector loops with arbitrary numbers of external $\bar\phi$ legs.
\end{itemize}

\item Note that the field $\bar\phi$ enters the action (\ref{s1.22}) only in
the dimensionless combination $(I\bar\phi)$. Then, by the dimensionality
reasoning, the most general form of the effective action depending on
$(I\bar\phi)$ should be of the following form
\be
\Gamma=\int d^4x[
f_1(I\bar\phi)I^2\square\bar\phi\square\bar\phi
+f_2(I\bar\phi)I^3\square\bar\phi
\partial_m\bar\phi\partial_m\bar\phi
+f_3(I\bar\phi)I^4(\partial_m\bar\phi\partial_m\bar\phi)^2]\,,
\label{s1.23}
\ee
where $f_1,\ f_2,\ f_3$ are some functions. The Feynman graph
computations should specify the unknown functions in (\ref{s1.23}).
\end{enumerate}

The property {\bf iv} implies that the effective action can be
written as \footnote{Note that the one-loop effective action
in the Euclidean space is given by $\Gamma=\frac 12\Tr\ln S''[\Phi]$ rather
then the Minkowski space expression $\Gamma=\frac i2\Tr\ln S''[\Phi]\,$.
Here $S''[\Phi]$ is the second functional derivative of the classical action.}
\be
\Gamma^{SYM}=\frac12\Tr\ln\frac{\delta^2 \tilde S}{\delta A_p(x)\delta
A_q(x')}\,,
\label{s1.23_}
\ee
where the action $\tilde S$ is the last line in (\ref{s1.22})
\be
\tilde S=\int d^4x\left[-\frac12 A_n\square A_n+ A_n G_m\partial_n A_m-
A_n G_n\partial_m A_m-\varepsilon_{mnrs}G_mA_n\partial_r
A_s\right].
\label{s1.24}
\ee
The second functional derivative in (\ref{s1.23_}) can be easily
calculated
\be
\frac{\delta^2\tilde S}{\delta A_p(x)\delta A_q(x')}=
-\delta_{pq}\square\delta^4(x-x')
+4G_{[q}\partial_{p]}\delta^4(x-x')
+2\varepsilon_{pqmn} G^m\partial^n\delta^4(x-x')\,.
\label{s1.25}
\ee
Substituting the expression (\ref{s1.25}) into (\ref{s1.23_}) we have
\begin{eqnarray}
\Gamma^{SYM}&=&\frac 12 \Tr\ln[
\delta_{pq}\delta^4(x-x')
+4G_{[p}\partial_{q]}\frac{1}\square\delta^4(x-x')
-2\varepsilon_{pqmn} G_m\frac{\partial_n}\square\delta^4(x-x')]
\nonumber\\&=&
\frac 12 \Tr\sum_{j=1}^\infty \frac{(-1)^{j+1}}j
[4G_{[p}\partial_{q]}\frac{1}\square\delta^4(x-x')
-2\varepsilon_{pqmn} G_m\frac{\partial_n}\square\delta^4(x-x')]^j\,.
\label{s1.26}
\end{eqnarray}
The sum in (\ref{s1.26}) is taken over the external legs $G_m$,
where the field $G_m$ depends on $\bar\phi$ according to the
definition (\ref{s1.14}).
Note that the expression (\ref{s1.26}) can be equivalently rewritten
in the form
\be
\Gamma^{SYM}=\frac12\Tr\sum^\infty_{j=1}
\frac{(-2)^{j+1}}{j}[X_{mnp}(x)\partial_n\frac1\square
\delta^4(x-x')]^j\,,
\label{s1.26_1}
\ee
where we have introduced the superfield
\be
X_{mnp}=G_m\delta_{np}-G_p\delta_{mn}
-G_q\varepsilon_{qpnm}\,.
\label{s1.26_2}
\ee
The representation of the action (\ref{s1.26_1}) in terms of Feynman diagrams
is given in the Appendix A.2 (Fig. 1a).

The propagators in (\ref{s1.26}) appear in the combination
$\partial_m{\square}^{-1}\delta^4(x-x')\,$. On the dimensionality
grounds, only the expressions like
\be
\left[\frac{\partial_m}\square\delta^4(x-x')\right]^2,\quad
\left[\frac{\partial_m}\square\delta^4(x-x')\right]^3,\quad
\left[\frac{\partial_m}\square\delta^4(x-x')\right]^4
\label{s1.26_}
\ee
are divergent, all higher powers of these expressions produce
finite contributions to the effective action. Therefore, only two-
three- and four-point diagrams make the divergent contributions to
the effective action (note that the external line is that of the field
$G_m$). We are interested solely in the divergent contributions to the
effective action, so we consider the calculations of two-,
three- and four-point functions separately.

According to eq. (\ref{s1.26}), the two-point function is defined by
\begin{eqnarray}
\Gamma_2^{SYM}&=&- \int d^4x_1 d^4x_2
[2G_{[q}(x_1)\partial_{p]}\frac1\square\delta^4(x_1-x_2)
+\varepsilon_{pqmn}G_m(x_1)\partial_n
\frac1\square\delta^4(x_1-x_2)]\nonumber\\&&
\qquad\times[
2G_{[p}(x_2)\partial_{q]}\frac1\square\delta^4(x_2-x_1)+
\varepsilon_{qprs}G_r(x_2)\partial_s\frac1\square\delta^4(x_2-x_1)]\,.
\label{s1.27}
\end{eqnarray}
Unfolding the product of square brackets in (\ref{s1.27}) and passing
to the momentum space
\be
G_m(x)=\int \frac{d^4p}{(2\pi)^4} e^{ip_nx_n}\tilde G_m(p)\,,\qquad
\delta^4(x-x')=\int \frac{d^4p}{(2\pi)^4} e^{ip_n(x-x')_n}\,,
\label{s1.28}
\ee
we find
\be
\Gamma_2^{SYM}=-2 \int \frac{d^4p}{(2\pi)^8}
\tilde G_m(p)\tilde G_r(-p)[2\delta_{ms}\delta_{nr}-2\delta_{mr}\delta_{ns}]
\int d^4k \frac{k_n(p+k)_s}{k^2(p+k)^2}\,.
\label{s1.29}
\ee
The divergent momentum integrals are calculated in the Appendix A.1 (eqs.
(\ref{a10}), (\ref{a11})). Further we shall consider only the divergent part
of the effective action (\ref{s1.29}) \footnote{In the dimensional regularization
scheme all divergencies of the effective action appear with the gamma-function factor
$\Gamma(\varepsilon)=\frac1\varepsilon-\gamma+O(\varepsilon)\,$. Here
$\varepsilon=2-d/2\,$, $d$ is the dimension of space-time, $\gamma$ is the
Euler constant. Therefore, all divergent contributions to the effective action
enter with the factor $1/\varepsilon$, $\varepsilon\to0\,$.}
\be
\Gamma_{2,div}^{SYM}=\frac{2\pi^2}{\varepsilon} \int \frac{d^4p}{(2\pi)^8}
[\tilde G_m(p) \frac13(p_mp_n+\frac{p^2\delta_{mn}}2)\tilde G_n(-p)
-\tilde G_m(p) p^2 \tilde G_m(-p)]\,.
\label{s1.30}
\ee
Switching back to the coordinate space and applying the relations
(\ref{s1.14}) we obtain:
\begin{eqnarray}
\Gamma_{2,div}^{SYM}=\frac1{16\pi^2\varepsilon}\int d^4x \ln(1+4I\bar\phi)
\square^2\ln(1+4I\bar\phi)\,.
\label{s1.31}
\end{eqnarray}

Consider now the computation of the divergent part of the three-point function
\begin{eqnarray}
\Gamma_3^{SYM}&=&\frac43\Tr[
(2G_{[r}(x)\partial_{s]}-\varepsilon_{rsmn}G_m(x)\partial_n)
\frac1\square\delta^4(x-x')]^3\nonumber\\
&=&\frac{4}3\int d^4x_1 d^4x_2 d^4x_3
[(G_t(x_1)\partial_u-G_u(x_1)\partial_t-\varepsilon_{tumn}G_m(x_1)\partial_n)
\frac1\square\delta^4(x_1-x_2)]\nonumber\\
&&\qquad\times
[(G_u(x_2)\partial_w-G_w(x_2)\partial_u-\varepsilon_{uwrs}G_r(x_2)\partial_s)
\frac1\square\delta^4(x_2-x_3)]\nonumber\\
&&\qquad\times
[(G_w(x_3)\partial_t-G_t(x_3)\partial_w-\varepsilon_{wtpq}G_p(x_3)\partial_q)
\frac1\square\delta^4(x_3-x_1)]\,.
\label{s1.32}
\end{eqnarray}
As in the previous case, we do the products of all brackets, pass to the momentum space
and regularize the divergent integrals according to eq. (\ref{a12}).
As a result, we arrive at the following expression for the
divergent part of the three-point Green function
\begin{eqnarray}
\Gamma_{3,div}^{SYM}&=&-\frac{1}{4\pi^2\varepsilon}
\int d^4xG_mG_m\partial_n G_n\nonumber\\&=&
-\frac{1}{4\pi^2\varepsilon}
\int d^4x\partial_m\ln(1+4I\bar\phi)\partial_m\ln(1+4I\bar\phi)
\square\ln(1+4I\bar\phi)\,.
\label{s1.33}
\end{eqnarray}
Here we made use of the definition (\ref{s1.14}).

The same machinery can be applied for computing the four-point Green function
\begin{eqnarray}
\Gamma_4^{SYM}&=&-2\Tr[
(2G_{[r}(x)\partial_{s]}-\varepsilon_{rsmn}G_m(x)\partial_n)
\frac1\square\delta^4(x-x')]^4
\nonumber\\&=&-2\int d^4x_1d^4x_2d^4x_3d^4x_4
[(G_p\partial_q-G_q\partial_p-\varepsilon_{p'qq'p} G_{p'}\partial_{q'})\frac1\square
\delta^4(x_1-x_2)]\nonumber\\&&\times
[(G_q\partial_m-G_m\partial_q-\varepsilon_{m'mn'q} G_{m'}\partial_{n'})\frac1\square
\delta^4(x_2-x_3)]\nonumber\\&&\times
[(G_m\partial_n-G_n\partial_m-\varepsilon_{r'ns'm} G_{r'}\partial_{s'})\frac1\square
\delta^4(x_3-x_4)]\nonumber\\&&\times
[(G_n\partial_p-G_p\partial_n-\varepsilon_{tpun} G_t\partial_u)\frac1\square
\delta^4(x_4-x_1)]\,.
\label{s1.34}
\end{eqnarray}
The divergent part of the action (\ref{s1.34}) is given by (after regularization
of momentum integrals in accord with  (\ref{a13}) and careful counting of the coefficients)
\be
\Gamma_{4,div}^{SYM}=-\frac{5}{16\pi^2\varepsilon}\int d^4x(G_m G_m)^2
=-\frac{5}{16\pi^2\varepsilon}\int d^4x
[\partial_m\ln(1+4I\bar\phi)\partial_m\ln(1+4I\bar\phi)]^2\,.
\label{s1.35}
\ee

Finally, we should put together the divergent contributions from two-, three,- and
four-point functions given by (\ref{s1.31}), (\ref{s1.33}) and (\ref{s1.35}). The result
is the total one-loop divergent contribution to the effective action in the deformed
$\cN=(1,1)$ SYM model
\begin{eqnarray}
\Gamma_{div}^{SYM}&=&\frac{1}{16\pi^2\varepsilon}\int d^4x
\ln(1+4I\bar\phi)\square^2\ln(1+4I\bar\phi)\nonumber\\&&
-\frac{1}{4\pi^2\varepsilon}\int d^4x
\partial_m\ln(1+4I\bar\phi)\partial_m\ln(1+4I\bar\phi)
\square\ln(1+4I\bar\phi)\nonumber\\&&
-\frac{5}{16\pi^2\varepsilon}\int d^4x
[\partial_m\ln(1+4I\bar\phi)\partial_m\ln(1+4I\bar\phi)]^2\,.
\label{s1.36}
\end{eqnarray}
The expression (\ref{s1.36}), modulo a total derivative under the integral,
can be equivalently rewritten as
\be
\Gamma_{div}^{SYM}=\frac1{\pi^2\varepsilon}\int d^4x \frac{I^2\square\bar\phi
\square\bar\phi}{(1+4I\bar\phi)^2}
-\frac6{\pi^2\varepsilon}\int d^4x \frac{
4I^3\square\bar\phi\partial_m\bar\phi\partial_m\bar\phi}{(1+4I\bar\phi)^3}\,.
\label{s1.37}
\ee
The action (\ref{s1.37}) is the complete divergent part of the effective
action in the deformed abelian $\cN=(1,1)$ gauge model. It matches with
the previously guessed structure (\ref{s1.23}).

At first sight, the model looks non-renormalizable, since the quantum
computations produce the terms (\ref{s1.37}) which are absent
in the classical action (\ref{s1.6}). Therefore, in order to make the model
renormalizable we are led to extend the classical action (\ref{s1.6}) by
the two extra terms
\be
c_1\int d^4x\frac{I^2
\square\bar\phi\square\bar\phi}{(1+4I\bar\phi)^2}
+c_2\int d^4x
\frac{4I^3\square\bar\phi\partial_m\bar\phi\partial_m\bar\phi}{
(1+4I\bar\phi)^3}
\label{s1.38}
\ee
with some coupling constants $c_1,\ c_2\,$.
However, both these terms can be
removed by shifting the scalar field in the classical action
\be
\phi\longrightarrow\phi-2c_1\frac{I^2\square\bar\phi}{(1+4I\bar\phi)^2}
-2c_2\frac{4I^3\partial_m\bar\phi\partial_m\bar\phi}{(1+4I\bar\phi)^3}\,,
\quad{\rm while}\quad \bar\phi\longrightarrow\bar\phi\,.
\label{s1.39}
\ee
Therefore, the $\cN =(1,0)$ gauge model is renormalizable in the sense that all
divergencies can be removed by the redefinition of the scalar field $\phi\,$.
Note that the redefinition of fields of the form (\ref{s1.39}) can be made
in the functional integral (\ref{s1.21}). Since the Jacobian of such a
change of functional variables equals unity, the terms (\ref{s1.38}), being
added to the classical action (\ref{s1.6}), do not make new contributions
to the effective action. In the language of Feynman diagrams this means
that the terms (\ref{s1.38}) generate new vertices for the scalar field.
But due to lacking of the propagator $\langle\bar\phi\bar\phi\rangle\,$,
no loops with such vertices can be constructed.

This situation is completely analogous to the $\cN=(1/2,0)$ SYM model considered
in \cite{Jack} where it was demonstrated that the quantum computations
in this model generate the divergent terms which are not present in
the classical action, but these extra divergencies can be removed by a simple
shift of the gaugino field (the lowest component in $\cN=(1,1)$ gauge multiplet).
In our case the divergencies can also be removed by the shift of lowest component
of $\cN=(1,1)$ gauge multiplet (scalar field).

To summarize, we have calculated the full divergent contribution
to the effective action in the deformed abelian $\cN=(1,1)$ gauge model. It can be
written in the form of two terms  (\ref{s1.37}). Both these terms can be
removed by the redefinition of classical field $\phi$ of the form
(\ref{s1.39}). Therefore the abelian deformed $\cN=(1,1)$ gauge model with the
action (\ref{s1.6}) is renormalizable. It is easy to see that the
divergent terms (\ref{s1.37}) vanish on the classical equations of
motion, therefore the S-matrix in this model is divergence-free and in
this sense one can say that the model under
consideration is finite.

\setcounter{equation}0
\section{Renormalizability of non-anticommutative
neutral hypermultiplet}

In this Section we prove the renormalizability (finiteness)
of the non-anticommutative model of a neutral hypermultiplet interacting
with an abelian gauge superfield. Firstly we consider the case when the gauge
superfield is treated as an external background and then the case of
general $\cN=(1,0)$ non-anticommutative model, with both gauge and
hypermultiplet superfields on equal footing.

Let us extend the non-anticommutative $U(1)$ gauge model (\ref{s1.6}) by adding the
hypermultiplet fields interacting with the vector multiplet. As pointed
out in \cite{ILZ2}, it is possible to consider here the adjoint and
fundamental representations of non-anticommutative U(1) group.
These theories are called the neutral and charged hypermultiplet models,
respectively. We will study further only the neutral hypermultiplet
model since it becomes free in the undeformed limit $I\to0$ similarly to the deformed
abelian supersymmetric gauge  model considered in Sect. 2.
The model of charged hypermultiplet
is essentially different since it retains a non-vanishing interaction in the limit $I\to0$ and the
considerations of quantum aspects of this model within the component field
formulation is a much more complicated problem. The problem of computing the
effective action in the charged hypermultiplet model will be treated elsewhere.

The classical action of the neutral hypermultiplet model
in the harmonic superspace \cite{HSS} is given by \cite{ILZ2}
\be
S_{hyp}=\int d\zeta du\, \tilde q^+(D^{++}q^++V^{++}\star q^+-q^+\star
V^{++})\,.
\label{s2.1}
\ee
Here $q^+$, $\tilde q^+$ are hypermultiplet superfields, $V^{++}$ is the
vector multiplet field, $D^{++}$ is the harmonic covariant derivative,
$d\zeta du$ is the integration measure of the analytic harmonic
superspace. For details of the harmonic superspace approach see, e.g.,
book \cite{Book}. The action (\ref{s2.1}) is invariant under
the following gauge transformations
\be
\delta\tilde q^+=[\tilde q^+,\Lambda]_\star,\qquad
\delta q^+=[q^+,\Lambda]_\star,\qquad
\delta V^{++}=D^{++}\Lambda+[V^{++},\Lambda]_\star
\label{s2.2}
\ee
with gauge parameter $\Lambda$ being analytic superfield.
It is obvious from (\ref{s2.1}) that the model under consideration becomes free
in the limit $I\to0\,$.

The component form of the action (\ref{s2.1}) (with the auxiliary fields
eliminated) was given in \cite{ILZ2}:
\begin{eqnarray}
S_{hyp}&=&\dint d^4x \left[
\frac12(1+4I\bar\phi)^2\partial_m f^{ak}\partial_m f_{ak}+
\frac i2(1+4I\bar\phi)\rho^{\alpha a}\partial_{\alpha\dot\alpha}
\chi^{\dot\alpha}_a\right.\nonumber\\&&\left.
+4iI\bar\Psi^{\dot\alpha}_k\rho^\alpha_a\partial_{\alpha\dot\alpha}
f^{ak}+2iI\rho^{\alpha a}A_m\partial_m\rho_{\alpha a}
+iI\rho^{\beta a}\rho_a^\alpha\partial_{(\alpha\dot\alpha}A^{\dot\alpha}_{\beta)}
\right].
\label{s2.4}
\end{eqnarray}
Here $f^{ak}$, $\rho^{\alpha a},\ \chi_a^{\dot\alpha}$ are physical scalar and
spinor fields of the hypermultiplet, $\bar\phi$,
$\bar\Psi_k^{\dot\alpha}$, $A_m$ are physical scalar, spinor and vector fields
of the vector multiplet. The indices $a,k$ running over $1,2$ are doublet indices of
two independent internal symmetry SU(2) groups.

Like in the case of the non-anticommutative gauge theory, we can
give a formal proof of renormalizability of the model (\ref{s2.4})
by applying the Wilsonian criterion \cite{Wilson}. According to eq.
(\ref{s1.10_}), we should ascribe the following asymmetrical scaling
dimensions to the component fields of the hypermultiplet, \be
[f^{ak}]=1,\qquad [\rho^{\alpha a}]=1,\qquad
[\chi_a^{\dot\alpha}]=2. \label{s2.4_1} \ee Using
eqs.(\ref{s1.10_2}), (\ref{s2.4_1}), it is easy to check that the
dimensions of all operators in the action (\ref{s2.4}) are just 4:
\be
\begin{array}c
[(1+4I\bar\phi)^2\partial_m f^{ak}\partial_m f_{ak}]=4,\qquad
[(1+4I\bar\phi)\rho^{\alpha a}\partial_{\alpha\dot\alpha}
\chi^{\dot\alpha}_a]=4,\\
{}[\bar\Psi^{\dot\alpha}_k\rho^\alpha_a\partial_{\alpha\dot\alpha}
f^{ak}]=4,\qquad
[\rho^{\alpha a}A_m\partial_m\rho_{\alpha a}]=4,\qquad
[\rho^{\beta
a}\rho_a^\alpha\partial_{(\alpha\dot\alpha}A^{\dot\alpha}_{\beta)}]=4.
\end{array}
\label{s2.4_2} \ee Therefore, the action (\ref{s2.4}) satisfies the
conditions of the Wilsonian approach, and the abelian neutral
hypermultiplet model is renormalizable. However the details of the
renormalization procedure require further analysis.

Now we are going to compute directly the divergent contributions
to the effective action of neutral hypermultiplet model. As a prelude,
let us comment on the structure of eq. (\ref{s2.4}).
\begin{enumerate}
\item We consider the fields of the gauge multiplet ($\bar\phi$,
$\bar\Psi_k^{\dot\alpha}$, $A_m$) as the external fields which are not
quantized. The quantum fields are physical fields of the hypermultiplet.
Note that the hypermultiplet fields enter the action
(\ref{s2.4}) only quadratically, therefore the effective action in this
model is automatically one-loop exact. This is also clear from
the form of the superfield action (\ref{s2.1}).
\item As proved in Appendix A.3, the terms in the second line of
(\ref{s2.4}) do not make any contribution to the quantum effective action since
the corresponding vertices appear without their conjugates, and so the
Feynman rules do not allow to compose any loop from such vertices.
Therefore for quantum calculations only first two terms in the action
(\ref{s2.4}) are really essential, and in what follows we can limit our consideration
just to these terms.
\item The first two terms in (\ref{s2.4}) depend only on the background
field $\bar\phi$. Therefore, the whole effective action is a functional
of the form (\ref{s1.23}) containing only $\bar\phi$-dependence.
\item It is easy to prove that the term
$\frac i2(1+4I\bar\phi)\rho^{\alpha a}\partial_{\alpha\dot\alpha}
\chi^{\dot\alpha}_a$, which is responsible for the fermionic loop, does not
make any non-trivial contribution to the effective action. Indeed, let us
consider a part of the one-loop effective action which is produced by this
fermionic loop
\begin{eqnarray}
\Gamma_{ferm}&=&-\Tr\ln\frac{\delta^2 S_{hyp}}{\delta\rho^{\alpha a}(x)
\delta\chi_b^{\dot\alpha}(x')}=
-\Tr\ln\left[\frac i2(1+4I\bar\phi)\partial_{\alpha\dot\alpha}\delta^4(x-x')
\delta_a^b\right]\nonumber\\&=&
-2\Tr\ln[\frac i2(1+4I\bar\phi)\delta^4(x-x')]
-2\Tr\ln[\partial_{\alpha\dot\alpha}\delta^4(x-x')]
\simeq0\,.
\label{s2.5}
\end{eqnarray}
Both terms in the second line of (\ref{s2.5}) make only trivial
contributions to the effective action and so can be discarded.
\end{enumerate}

Taking into account these remarks, the effective action in the hypermultiplet model
(\ref{s2.4}) is defined by
\be
\Gamma^{hyp}=\frac12\Tr\ln\frac{\delta^2}{\delta f^{ak}(x)\delta f_{a'k'}(x')}
\left[\frac12\int d^4x (1+4I\bar\phi)^2\partial_m
f^{ak}\partial_m f_{ak} \right].
\label{s2.6}
\ee
Calculating the functional derivative in (\ref{s2.6}) and performing some
further manipulations, we find
\begin{eqnarray}
\Gamma^{hyp}&=&2\Tr\ln\left[(1+4I\bar\phi)^2\square\delta^4(x-x')
+\partial_m(1+4I\bar\phi)^2\partial_m\delta^4(x-x') \right]
\nonumber\\&=&
2\Tr\ln\left[(1+4I\bar\phi)^2\square\delta^4(x-x') \right]
\nonumber\\&&
+2\Tr\ln\left[\delta^4(x-x')+2\frac1\square G_m(x)\partial_m\delta^4(x-x')
 \right].
\label{s2.7}
\end{eqnarray}
The second line of (\ref{s2.7}) does not contribute to the effective
action since it is proportional to $\delta^4(0)$, which is zero
in the dimensional regularization scheme. Therefore,
making a series expansion of the expression in the last line of (\ref{s2.7}),
we obtain the following formal answer for the effective action,
\be
\Gamma^{hyp}=2\Tr\sum\limits_{n=1}^\infty\frac{(-1)^{n+1}}{n}
\left[\frac2\square G_m(x)\partial_m\delta^4(x-x')\right]^n.
\label{s2.8}
\ee
Eq. (\ref{s2.8}) is the starting point of the perturbative
calculation of the one-loop effective action in the neutral
hypermultiplet model. Resorting to the dimensional arguments, like in the
gauge model considered in Sect. 2, it is easy to show that
only two-, three-, and four-point functions
are divergent since they contain the momentum integrals corresponding to the
expressions (\ref{s1.26_}). As far as we are interested only in the divergent part
of the effective action, we will consider the computation of
two-, three-, and four-point diagrams separately.

According to eq. (\ref{s2.8}), the two-point function is defined by the
expression
\be
\Gamma^{hyp}_2=-4\int d^4x_1d^4x_2 G_m(x_1) G_n(x_2)
\frac{\partial_m}{\square}\delta^4(x_1-x_2)
\frac{\partial_n}{\square}\delta^4(x_2-x_1)\,.
\label{s2.9}
\ee
Passing to the momentum space by the standard rules (\ref{s1.28}),
we obtain
\be
\Gamma_2^{hyp}=-4\int\frac{d^4p}{(2\pi)^8}\tilde G_m(p)\tilde G_n(-p)
\int d^4k\frac{k_m(p+k)_n}{k^2(p+k)^2}\,.
\label{s2.10}
\ee
The divergent momentum integral was calculated in the Appendix A.1 (eq.
(\ref{a10})). Here we need only the divergent part of this integral which
reads
\be
\Gamma_{2,div}^{hyp}=\frac{2\pi^2}{3\varepsilon}
\int\frac{d^4p}{(2\pi)^8}\tilde G_m(p)\tilde G_n(-p)\left(
p_m p_n+\frac{\delta_{mn}}2p^2\right).
\label{s2.11}
\ee
Switching back to the configuration space, we obtain the divergent
two-point contribution to the effective action
\begin{eqnarray}
\Gamma_{2,div}^{hyp}&=&\frac2{3\varepsilon 16\pi^2}
\int d^4x[G_m(x)\partial_m\partial_n G_n(x)+\frac12 G_m(x)\square G_m(x)]
\nonumber\\&
=&-\frac1{16\pi^2\varepsilon}\int d^4x\ln(1+4I\bar\phi(x))\square^2
\ln(1+4I\bar\phi(x))\,.
\label{s2.12}
\end{eqnarray}
Up to the sign, the expression (\ref{s2.12}) is equal to the divergence
of two-point function (\ref{s1.31}) in the gauge model.

Consider now the three-point function
\begin{eqnarray}
\Gamma_3^{hyp}&=&\frac{16}3\int d^4x_1d^4x_2d^4x_3 G_m(x_1)G_n(x_2)G_p(x_3)
\nonumber\\&&\times
\frac{\partial_m}{\square}\delta^4(x_1-x_2)
\frac{\partial_n}{\square}\delta^4(x_2-x_3)
\frac{\partial_p}{\square}\delta^4(x_3-x_1)\,.
\label{s2.13}
\end{eqnarray}
Passing to the momentum space, we obtain
\begin{eqnarray}
\Gamma_3^{hyp}&=&\frac{16}3\int\frac{d^4p_1d^4p_2d^4p_3}{(2\pi)^{12}}
\tilde G_m(p_1)\tilde G_n(p_2)\tilde G_p(p_3)\delta^4(p_1+p_2+p_3)
\nonumber\\&&\times
\int d^4k\frac{k_m(k-p_2)_n(k+p_1)_p}{k^2(k-p_2)^2(k+p_1)^2}\,.
\label{s2.14}
\end{eqnarray}
The divergent part of the momentum integral was calculated in the Appendix A.1,
eq. (\ref{a12}). Using this result, we find
\begin{eqnarray}
\Gamma_{3,div}^{hyp}&=&\frac{4\pi^2}{9\varepsilon}
\int\frac{d^4p_1d^4p_2d^4p_3}{(2\pi)^{12}}\delta^4(p_1+p_2+p_3)
[\tilde G_m(p_1)\tilde G_m(p_2)\tilde G_p(p_3)(2p_1+p_2)_p\nonumber\\&&
-\tilde G_m(p_1)\tilde G_n(p_2)\tilde G_m(p_3)(p_1+2p_2)_n
-\tilde G_m(p_1)\tilde G_n(p_2)\tilde G_n(p_3)(p_1-p_2)_m]\nonumber\\
&=&-\frac{1}{8\pi^2\varepsilon}\int d^4x(\partial_m G_m)G_nG_n\,.
\label{s2.15}
\end{eqnarray}
Thus the contribution to the divergent part of the effective
action from this term reads
\be
\Gamma_{3,div}^{hyp}=-\frac1{8\pi^2\varepsilon}
\int d^4x\square\ln(1+4I\bar\phi)
\partial_n\ln(1+4I\bar\phi)\partial_n\ln(1+4I\bar\phi)\,.
\label{s2.16}
\ee

Finally, let us consider the computation of four-point function
\begin{eqnarray}
\Gamma_4^{hyp}&=&-8\int d^4x_1 d^4x_2d^4x_3d^4x_4
G_m(x_1)G_n(x_2)G_p(x_3)G_r(x_4)\nonumber\\&&\times
\frac{\partial_m}\square\delta^4(x_1-x_2)
\frac{\partial_n}\square\delta^4(x_2-x_3)
\frac{\partial_p}\square\delta^4(x_3-x_4)
\frac{\partial_r}\square\delta^4(x_4-x_1)\nonumber\\
&=&-8\int\frac{d^4p_1\ldots d^4p_4}{(2\pi^{16})}
\tilde G_m(p_1)\tilde G_n(p_2)\tilde G_p(p_3)\tilde G_r(p_4)
\delta^4(p_1+p_2+p_3+p_4)\nonumber\\&&\times
\int d^4k\frac{k_m(k-p_2)_n(k+p_1+p_4)_p(p_1+k)_r}{
k^2(k-p_2)^2(k+p_1+p_4)^2(p_1+k)^2}\,.
\label{s2.17}
\end{eqnarray}
Substituting the expression (\ref{a13}) for the divergent momentum integral,
we obtain the following expression for the divergent part of four-point function
\begin{eqnarray}
\Gamma_{4,div}^{hyp}&=&-\frac{\pi^2}{3\varepsilon}
\int\frac{d^4p_1\ldots d^4p_4}{(2\pi)^{16}}\delta^4(p_1+p_2+p_3+p_4)
3\tilde G_m(p_1)\tilde G_m(p_2)\tilde G_n(p_3)\tilde G_n(p_4)\nonumber\\
&=&-\frac{1}{16\pi^2\varepsilon}\int d^4x[G_m(x)G_m(x)]^2\,.
\label{s2.18}
\end{eqnarray}
As a result, the corresponding contribution to the effective
action is given by
\be
\Gamma_{4,div}^{hyp}=-\frac{1}{16\pi^2\varepsilon}\int d^4x
[\partial_m\ln(1+4I\bar\phi)
\partial_m\ln(1+4I\bar\phi)]^2\,.
\label{s2.19}
\ee

Now we sum up all the divergent contributions to the effective action
 found above, i.e. (\ref{s2.12}), (\ref{s2.16}) and (\ref{s2.19})
\begin{eqnarray}
\Gamma_{div}^{hyp}&=&-\frac1{16\pi^2\varepsilon}\int d^4x
\ln(1+4I\bar\phi)\square^2\ln(1+4I\bar\phi)
\nonumber\\&&
-\frac1{8\pi^2\varepsilon}\int d^4x\square\ln(1+4I\bar\phi)
\partial_n\ln(1+4I\bar\phi)\partial_n\ln(1+4I\bar\phi)
\nonumber\\&&
-\frac1{16\pi^2\varepsilon}\int d^4x[\partial_m\ln(1+4I\bar\phi)
\partial_m\ln(1+4I\bar\phi)]^2\,.
\label{s2.20}
\end{eqnarray}
After some work all three terms in the effective action (\ref{s2.20}) can
be shown to reduce to the following simple expression
\be
\Gamma_{div}^{hyp}=-\frac{1}{\pi^2\varepsilon}\int d^4x\frac{I^2
\square\bar\phi\square\bar\phi}{(1+4I\bar\phi)^2}\,.
\label{s2.21}
\ee
The expression (\ref{s2.21}) represents the complete divergent contribution
to the effective action in the deformed model of hypermultiplet interacting
with the abelian gauge multiplet. Once again, eq. (\ref{s2.21}) matches with
the general ansatz (\ref{s1.23}).

As in the deformed gauge model, in order to cancel  the divergent term (\ref{s2.21})
one should add the corresponding counterterm to the classical action of the
gauge model (\ref{s1.6}).
In other words, to make the model renormalizable we
should add to the classical action (\ref{s1.6}) the expression
\be
c_1\int d^4x\frac{I^2
\square\bar\phi\square\bar\phi}{(1+4I\bar\phi)^2}\,,
\label{s2.22}
\ee
where $c_1$ is some constant.
Remarkably, in a close similarity to the consideration in the gauge model,
the expression (\ref{s2.22}) can be completely absorbed into a redefinition
of another scalar field of the gauge multiplet in the classical action of
the gauge model:
\be
\phi\longrightarrow
\phi-2c_1\frac{I^2\square\bar\phi}{(1+4I\bar\phi)^2}\,,
\quad {\rm while}\quad \bar\phi\longrightarrow\bar\phi\,.
\label{s2.23}
\ee
Therefore, the appearance of such a divergent term does not spoil
the renormalizability of the theory in the sense that it
can be removed by redefining the scalar
field $\phi$. On the quantum level, the term (\ref{s2.22}) does not make any
contribution to the effective action of the model since we can perform the change
of functional variables (\ref{s2.23}) in the functional integral.

Let us now consider the general abelian $\cN=(1,0)$ non-anticommutative model of
gauge superfield field interacting with the hypermultiplet matter.
It is described by the classical action
\be
S=S_{SYM}+S_{hyp},
\label{s2.24}
\ee
where $S_{SYM}$ and $S_{hyp}$ are given by (\ref{s1.6}) and (\ref{s2.4}). Using
the Feynman rules developed in the Appendices A.2, A.3, it is easy to demonstrate
that the total divergent contribution in the model (\ref{s2.24}) is a sum of
divergencies of each model (\ref{s1.37}) and (\ref{s2.21})
\be
\Gamma_{div}=\Gamma_{div}^{SYM}+\Gamma_{div}^{hyp}=
-\frac6{\pi^2\varepsilon}\int d^4x \frac{
4I^3\square\bar\phi\partial_m\bar\phi\partial_m\bar\phi}{(1+4I\bar\phi)^3}\,.
\label{s2.25}
\ee
The divergent term (\ref{s2.25}) can also be removed by a shift of the
scalar field $\phi$
\be
\phi\longrightarrow\phi
-2c_2\frac{4I^3\partial_m\bar\phi\partial_m\bar\phi}{(1+4I\bar\phi)^3}\,,
\label{s2.26}
\ee
where $c_2=-6/(\pi^2\varepsilon)$ in this case. After the field redefinition
(\ref{s2.26}) the effective action of general abelian $\cN=(1,0)$
non-anticommutative model is divergence-free. Since the divergent term (\ref{s2.25})
vanishes on the classical equations of motion, the S-matrix in this
models is finite.

\setcounter{equation}0
\section{Seiberg-Witten map and renormalizability}

The Seiberg-Witten map for $\cN=(1,0)$ gauge model (\ref{s1.6}) was found in
\cite{FILSZ}. After the redefinition of fields
\begin{eqnarray}
\varphi&=&\phi+4I(1+4I\bar\phi)^{-1}[
A_mA_m+4I^2(\partial_m\bar\phi)^2],\nonumber\\
a_m&=&(1+4I\bar\phi)^{-1}A_m,\qquad
\bar\psi^k_{\dot\alpha}=(1+4I\bar\phi)^{-1}\bar\Psi^k_{\dot\alpha},
\nonumber\\
\psi^k_\alpha&=&\Psi^k_\alpha+
4I(1+4I\bar\phi)^{-1}A_{\alpha\dot\alpha}\bar\Psi^{\dot\alpha k},
\nonumber\\
d^{kl}&=&(1+4I\bar\phi)^{-1}[{\cal D}^{kl}+8I(1+4I\bar\phi)^{-1}
\bar\Psi^k_{\dot\alpha}\bar\Psi^{\dot\alpha l}]
\label{s3.1}
\end{eqnarray}
the action (\ref{s1.6}) simplifies drastically to
\begin{eqnarray}
S_{SYM}&=&\int d^4x (
-\frac12\varphi\square \bar\phi-i\psi_k^\alpha\partial_{\alpha\dot\alpha}
\bar\psi^{\dot\alpha k}+\frac14d^{kl}d_{kl})
\nonumber\\&&+
\frac14\int d^4x(1+4I\bar\phi)^2
(f_{mn}f_{mn}+\frac12\varepsilon_{mnrs}f_{mn}f_{rs})\,.
\label{s3.2}
\end{eqnarray}
Here $f_{mn}=\partial_m a_n-\partial_n a_m$. Note that the spinor and
auxiliary fields are free, while the interaction between the vector field and
the scalar $\bar\phi$ in the second line of (\ref{s3.2}) is still essential.

The action (\ref{s3.2}) is invariant under the abelian gauge
transformations
\be
\delta a_m=\partial_m \lambda
\label{s3.3}
\ee
with $\lambda$ being the gauge parameter.
Therefore we use standard Lorentz gauge fixing
\be
\partial_m a_m=0\,.
\label{s3.4}
\ee
Following the Faddeev-Popov procedure for constructing the functional integral,
we introduce the gauge fixing function
\be
\chi=\partial_m a_m\,,
\label{s3.5}
\ee
which transforms as $\delta \chi=\square\lambda\,.$ Therefore the ghost fields do
not interact with other fields and completely decouple. The ghost action
is given again by eq. (\ref{s1.16}).
The generating functional for Green's functions is now given by
\footnote{Note that the Jacobian of the change of functional variables
(\ref{s3.1}) is unity since this redefinition of fields is local.}
\be
Z[J]=\int {\cal D}(\varphi,\bar\phi,\psi,\bar\psi, a_m,b,c)
\delta(\chi-\partial_m a^m)
e^{-\frac12(S_{SYM}+S_{FP}+S_J)}\,,
\label{s3.6}
\ee
where
\be
S_J=\int d^4x[\varphi J_\varphi+\bar\phi J_{\bar\phi}
+\psi^i_\alpha (J_{\psi})_i^\alpha+
\bar\psi_{i\dot\alpha} (J_{\bar\psi})^{i\dot\alpha}+ a_m (J_a)_m]\,.
\label{s3.6.1}
\ee
To represent the delta-function in
the Gaussian form, we average the equation (\ref{s3.6}) with the functional
factor (\ref{s1.19}). As a result we obtain the gauge fixing action in the form
\be
S_{gf}=\frac\alpha2\int d^4x(1+4I\bar\phi)^2\partial_m a_m
\partial_n a_n\,.
\label{s3.7}
\ee
For simplicity we choose the gauge fixing parameter $\alpha$ to be unity,
$\alpha=1\,$. As a result, the generating functional (\ref{s3.6}) reads
\be
Z[J]=\int{\cal D}(\varphi,\bar\phi,\psi,\bar\psi, a_m, b,c)
e^{-\frac12(S_{\rm tot}+S_{FP}+S_J)}\,,
\label{s3.8}
\ee
where
\be
S_{\rm tot} =S_{SYM}+S_{gf}
=\int d^4x (
-\frac12\varphi\square \bar\phi-i\psi_k^\alpha\partial_{\alpha\dot\alpha}
\bar\psi^{\dot\alpha k}+\frac14d^{kl}d_{kl})
+S_a\,,
\label{s3.9}
\ee
and
\be
S_a=\frac12\int d^4x(1+4I\bar\phi)^2
(\partial_m a_m \partial_n a_n+
\partial_m a_n\partial_m a_n-
\partial_m a_n\partial_n a_m
+\varepsilon_{mnrs}\partial_m a_n\partial_r a_s)\,.
\label{s3.10}
\ee

It is evident that the scalar and spinor fields as well as the ghosts do not
contribute to the effective action. The only contribution comes from the
part (\ref{s3.10}), namely
\begin{eqnarray}
\Gamma^{SYM}&=&\frac12\Tr\ln\frac{\delta^2 S_a}{\delta a_p(x)\delta a_q(x')}
\nonumber\\
&=&\frac12\,\Tr\ln[
\delta_{pq}\square\delta^4(x-x')+2\delta_{pq}G_m\partial_m\delta^4(x-x')
\nonumber\\&&
+\,4 G_{[p}\partial_{q]}\delta^4(x-x')-2\varepsilon_{pqmn}G_m
\partial_n\delta^4(x-x')
]\,.
\label{s3.11}
\end{eqnarray}
The field $G_m(x)$ was defined in eq. (\ref{s1.14}).
The expression (\ref{s3.11}) is the starting point for perturbative
calculations of one-loop effective action in the $\cN=(1,0)$
non-an\-ti\-com\-mu\-ta\-ti\-ve SYM model. Note that it resembles the first line
of eq.(\ref{s1.26}), except for the term
$2\delta_{pq}G_m\partial_m\delta^4(x-x')\,$. Therefore the further computations
are very similar to ones given in Sect. 2. As usual, only two-, three- and
four-point diagrams are divergent. The two-point function is given by
\begin{eqnarray}
\Gamma_2^{SYM}&=&-\int d^4x_1 d^4x_2
[\delta_{pq}G_m(x_1)\partial_m\frac{1}{\square}\delta^4(x_1-x_2)+
2G_{[p}(x_1)\partial_{q]}\frac1\square\delta^4(x_1-x_2)
\nonumber\\&&
+\,\varepsilon_{qpmn}G_m(x_1)\partial_n
\frac1\square\delta^4(x_1-x_2)]
[\delta_{qp}G_n(x_2)\partial_n\frac{1}{\square}\delta^4(x_2-x_1)
\nonumber\\&&+
\,2G_{[q}(x_2)\partial_{p]}\frac1\square\delta^4(x_2-x_1)+
\varepsilon_{pqrs}G_r(x_2)\partial_s\frac1\square\delta^4(x_2-x_1)]\,.
\label{s3.12}
\end{eqnarray}
To proceed, we pass to the momentum space and
compute the divergent momentum integrals according to eqs.
(\ref{a10}), (\ref{a11}). As a result we find that the two-point function
(\ref{s3.12}) has no divergent contributions, i.e.
\be
\Gamma^{SYM}_{2,div}=0\,.
\label{s3.13}
\ee
The absence of divergencies here is owing to the term
$2\delta_{pq}G_m\partial_m\delta^4(x-x')$ in (\ref{s3.11}) and (\ref{s3.12}).
It gives the contribution which
exactly cancels the expression (\ref{s1.31}) obtained by similar
calculations without this term.

The three- and four-point functions are defined by the following formal
expressions:
\begin{eqnarray}
\Gamma_3^{SYM}&=&\frac43\Tr[(\delta_{pq}G_m(x)\partial_m
+2G_{[p}(x)\partial_{q]}-\varepsilon_{pqmn}G_m(x)\partial_n)
\frac1\square\delta^4(x-x')]^3,
\label{s3.14}\\
\Gamma_4^{SYM}&=&-2\Tr[(\delta_{pq}G_m(x)\partial_m
+2G_{[p}(x)\partial_{q]}-\varepsilon_{pqmn}G_m(x)\partial_n)
\frac1\square\delta^4(x-x')]^4.
\label{s3.15}
\end{eqnarray}
The further computations are very similar to those in Sect.2, but with taking into
account the term $2\delta_{pq}G_m\partial_m\delta^4(x-x')\,$. After carefully
tracking the coefficients during the computations, we find that the three- and
four-point functions also have no divergences,
\be
\Gamma^{SYM}_{3,div}=0\,,\qquad
\Gamma^{SYM}_{4,div}=0\,.
\label{s3.16}
\ee
As a result, we conclude that the abelian $\cN=(1,0)$ non-anticommutative gauge
model (\ref{s3.2}) is completely finite, thus
\be
\Gamma^{SYM}_{div}=0
\label{s3.17}
\ee
without the necessity to perform any field redefinition such as (\ref{s1.39}).

The absence of divergencies in the abelian $\cN=(1,0)$ non-anticommutative gauge
model confirms the results of Sect.2, where these calculations were performed
without the use of Seiberg-Witten map (\ref{s3.1}). This is a consequence of
the fact that the considered model has a very specific interaction due to the
non-anticommutativity.

One more important comment to be added is as follows. The abelian $\cN=(1,0)$
non-anticommutative  gauge model is described by the classical actions (\ref{s1.6})
or (\ref{s3.2}) which are related to each other through the Seiberg-Witten map
(\ref{s3.1}). It is obvious that the Jacobian of such a change of functional
variables (\ref{s3.1}) is unity (in the sense of dimensional regularization).
Therefore the effective actions in these two models should also be related by
the Seiberg-Witten map. As for the divergent part, we observe that it is trivial
for both models (\ref{s1.6}) and (\ref{s3.2}), since it can be removed by
the shift (\ref{s1.39}) of the scalar field $\phi\,$. Note that this
explains the appearance of only two out of three possible divergent
terms (\ref{s1.23}). Indeed, if the third term proportional to
$I^4\int d^4x f_3(I\bar\phi)(\partial_m\bar\phi\partial_m\bar\phi)^2$ appeared in
the divergent part of the effective action, it could not be removed by any shift
of the scalar field $\phi\,$, which would mean the presence of
a nontrivial divergence in the model. However, we have seen in this Section
that the effective action in $\cN=(1,0)$ non-anticommutative gauge theory is
finite.

Let us consider also the general model of an abelian $\cN=(1,0)$ non-anticommutative
gauge superfield interacting with a neutral hypermultiplet. It is described by the sum
of the classical actions (\ref{s1.6}) and (\ref{s2.4}). In \cite{ILZ2} it was
shown that, after the appropriate redefinition of fields (Seiberg-Witten map), the
action of this model is given by
\begin{eqnarray}
S&=&S_0+S_{1}\,,\label{s3.19}\\
S_0&=&\int d^4x[
-\frac12\hat\varphi\square\bar\phi+
\frac12\partial_m \hat f^{ak}\partial_m \hat f_{ak}-
i\hat\psi^\alpha_k\partial_{\alpha\dot\alpha}
\bar\psi^{\dot\alpha k}+\frac i2\hat \rho^{\alpha a}
\partial_{\alpha\dot\alpha} \hat \chi^{\dot\alpha}_{a}+\frac14 d_{kl}d^{kl}],
\label{s3.20}\\
S_1&=&\int d^4x[\frac14(1+4I\bar\phi)^2(f_{mn}f_{mn}+\frac12\varepsilon_{mnrs}
f_{mn}f_{rs})+I(1+4I\bar\phi)^{-1}
\hat\rho^{\beta a}\hat\rho^\alpha_{a}f_{\alpha\beta}]\,,
\label{s3.21}
\end{eqnarray}
where $f_{\alpha\beta}=i(\partial_{\alpha\dot\alpha} a^{\dot\alpha}_\beta
+\partial_{\beta\dot\alpha}a_\alpha^{\dot\alpha})$ is one of two
self-dual parts of the Maxwell field strength $f_{mn}\,$.
The corresponding Seiberg-Witten map reads
\begin{eqnarray}
\hat f^{ak}&=&(1+4I\bar\phi)f^{ak}\,,\qquad
\hat\rho^{\alpha a}=(1+4I\bar\phi)\rho^{\alpha a}\,,\nonumber\\
\hat\chi^{\dot\alpha a}&=&\chi^{\dot\alpha a}+
4Ia^{\alpha\dot\alpha}\rho_\alpha^a
-8I\bar\psi^{\dot\alpha k}f_k^a\,,\nonumber\\
\hat\varphi&=&\varphi-4I(1+4I\bar\phi)(f^{ak}f_{ak})\,,\nonumber\\
\hat\psi_k^\alpha&=
&\psi^\alpha_k-4I(1+4I\bar\phi)(\rho^{\alpha a}f_{ak})\,.
\label{s3.22}
\end{eqnarray}

Note that the action $S_0$ (\ref{s3.20}) is free and it does not contribute
to the effective action. It is easy to demonstrate that the last term
in (\ref{s3.21}) also does not give rise to any quantum correction
since it is impossible to form any loop with such interactions.
The only non-trivial contribution to the effective action
comes from the first term in (\ref{s3.21}),
\be
\int d^4x\frac14(1+4I\bar\phi)^2(f_{mn}f_{mn}+\frac12\varepsilon_{mnrs}
f_{mn}f_{rs})\,.
\label{s3.22.1}
\ee
This expression just coincides with the one present in the
gauge theory action (\ref{s3.2}). Thus the quantum computations tell us
once again that the general abelian $\cN=(1,0)$ non-anticommutative model
is finite
\be
\Gamma_{div}=0\,.
\label{s3.23}
\ee
This result agrees with the one of Sect.3, modulo some divergent redefinition
(\ref{s2.26}) of the scalar field $\phi\,$.

To summarize, the use of the Seiberg-Witten map in the models under
consideration makes it possible to avoid the divergent
expressions in the effective action from the very beginning. Otherwise,
such expressions appear but they are removable by some divergent
redefinition of the scalar field $\phi\,$.

\section{Concluding remarks}

In this paper we addressed the problem of renormalizability of two supersymmetric models
with the nilpotent singlet deformation $\cN =(1,1) \rightarrow \cN = (1,0)$:
the model of abelian $\cN=(1,1)$ gauge vector multiplet, as well as the
model of abelian vector multiplet interacting with a neutral
hypermultiplet. Our main conclusion is that both these models are finite.

The consideration is based on component field computations of all divergent Feynman
graphs and their regularization. We observe the following common features peculiar to
both considered models.
\begin{enumerate}
\item The renormalizability of these models can be established
by ascribing non-standard scaling dimensions to the component fields
and then resorting to the general Wilsonian argument.
\item The effective action is defined only by one-loop contributions. The vertices
corresponding to the new interaction induced by the non-anticommutativity have a very
specific structure that ensures the absence of higher-loop contributions to the
effective action.
\item The analysis of the Feynman rules in the models shows that the effective
action depends only on the field $\bar\phi$ (but not on $\phi$). In other
words, only the field $\bar\phi$ can appear as the external legs while other
fields can propagate only inside the loop.
\item The diagrams with fermionic fields inside the loop do not contribute to
the effective action (more precisely, these diagrams give an infinite
contribution which is commonly discarded within the dimensional regularization
scheme). There are only two types of non-trivial diagrams: with the vector field inside
the loop in the gauge model and with the scalar fields inside the loop
in the hypermultiplet model.
\item The divergent diagrams carry only two, three or four external
legs. Any diagrams with more external legs are convergent.
\item The total divergent contribution to the effective action can be written in
the form of two terms (\ref{s1.38}) or one term (\ref{s2.22}), which are absent
in the original classical actions
of the gauge model or the gauge-hypermultiplet model, respectively. However, these
divergences can be eliminated by simple
redefinitions of the scalar field $\phi$ as in (\ref{s1.39}) or (\ref{s2.23}).
Since such a change of fields can be performed in the functional integral
defining the effective action (the Jacobian of such a change is unity),
we conclude that all divergencies can be eliminated by such redefinitions.
An important property is that the coupling constant (deformation parameter) $I$
is not renormalized, so its $\beta$-function is equal to zero.
Note also that the divergent terms (\ref{s1.38},\ref{s2.22}) vanish on the
classical equations of motion, therefore the S-matrix is divergence-free.
In this sense the considered models are finite.
\item In the $\cN=(1,0)$ non-anticommutative gauge models, both with
and without the hypermultiplet, there exists a Seiberg-Witten
map which essentially simplifies the classical actions of these theories.
It is an amazing feature
of the considered models that in terms of the new fields (after performing
the Seiberg-Witten map)
the quantum effective action is completely free of divergencies. This emphasizes
the ``unphysical'' nature of the divergent terms which appear when using
the original fields (before performing the Seiberg-Witten map).
\end{enumerate}

All these properties look rather strange since they are not featured
by conventional field models. However, these peculiarities are explained by the
fact that the considered models become free when the
non-anticommutativity is turned off. In this connection,
it seems important to study the renormalizability and the problem of effective
action in various $\cN=(1,0)$ non-anticommutative models which remain
interacting in the undeformed limit. One of the simplest theories of this kind is
a charged hypermultiplet interacting with an external abelian
gauge superfield. As is well known
(see e.g. \cite{Grisaru,Buch}), the low-energy effective action of the undeformed
charged hypermultiplet model is described by the holomorphic potential.
Therefore, it is very interesting to find the analogous contributions to
the effective action in the corresponding non-anticommutative model.
Note that the similar problems for $\cN=(1/2,0)$ Wess-Zumino and gauge models were
successfully solved in the works \cite{0307091,B}. Also it would be useful to
investigate the next-to-leading corrections. In conventional (undeformed)
$\cN=2,4$ gauge models such corrections form the non-holomorphic effective
potential having rather universal form \cite{non-hol}.
It would be interesting to clarify the structure of next-to-leading
corrections in the non-anticommutative theories.

Apart from the feature that the considered field theories look like the ``toy''
models since they become free in the undeformed limit, the proof of their
renormalizability is an important first step in attacking the issue of
renormalizability of deformed general non-abelian $\cN=(1,0)$ gauge theories.
Indeed, these models appear as a U(1) part of general non-abelian $\cN=(1,0)$ gauge
theories. The
renormalizability in the U(1) sector is necessary (but of course not sufficient)
for the whole non-abelian theory to be renormalizable (see, e.g., the
analysis of renormalizability of the $\cN=(1/2,0)$ SYM model in \cite{Jack,Grisaru2}).
However, the non-abelian generalization of our results is a very non-trivial
task, since for the time being the non-abelian deformed models
are insufficiently studied even at the classical level \cite{FILSZ,ILZ2}.

Another possible direction of extending our results is related to the issue
of renormalizability of non-anticommutative $\cN=(1,1)$ models with non-singlet
deformations considered e.g. in \cite{Araki1,CILQ}.

\section*{Acknowledgements}
I.B.S. is grateful to N.G. Pletnev for useful discussions. The work of I.L.B,
and I.B.S., as well as of E.A.I. and B.M.Z., was supported by the RFBR
grants No 03-02-16193 and No 03-02-17440, respectively.
The work of I.L.B., E.A.I., O.L. and B.M.Z. was also
supported by the DFG grant No 436 RUS 113/669. I.L.B., E.A.I. and B.M.Z.
acknowledge
a partial support from the
joint RFBR-DFG grant No 04-02-04002. O.L. acknowledges support from
the DFG grant Le-838/7 and I.L.B. from the grant INTAS-03-51-6346 and
a grant for LRSS, project No 1252.2003.2.
The work of I.B.S was partially supported by the joint DAAD -- Mikhail Lomonosov
Program (project No 71627) and by a grant for LRSS, project No 1743.2003.2.
E.A.I. and B.M.Z. acknowledge support from a grant of the Heisenberg-Landau program
and the NATO grant PST.GLG.980302. I.L.B. is grateful to Max Planck
Institut f\" ur Physik (Munich)
where the paper has been finalized and to Prof. D. L\"ust for kind hospitality.
I.B.S. thanks the Institut f\" ur Theoretische Physik, Universit\" at Hannover,
and E.A.I. thanks Laboratoire de Physique, ENS-Lyon, for the kind
hospitality extended to them at the final stage of this work.

\def\theequation{A.\arabic{equation}}
\setcounter{equation}0
\appendix
\section{Appendices}
\subsection{Divergent momentum integrals}
All divergent momentum (loop) integrals in quantum field theory can be
calculated using the dimensional regularization. For example, there is the
list of standard formulas \cite{Frampton}
\footnote{Note that we work in the Euclidean space, therefore our expressions
differ from those given in \cite{Frampton} by the imaginary unit factor.}
\begin{eqnarray}
\int \frac{d^dk}{(k^2+2kQ+M^2)^\alpha}&=&\frac{\pi^{d/2}}{
\Gamma(\alpha)(M^2-Q^2)^{\alpha-d/2}}\Gamma\left(\alpha-\frac d2\right)
\label{a1},\\
\int \frac{d^dk\, k_m}{(k^2+2kQ+M^2)^\alpha}&=&\frac{-Q_m\pi^{d/2}}{
\Gamma(\alpha)(M^2-Q^2)^{\alpha-d/2}}\Gamma\left(\alpha-\frac d2\right)
\label{a2},\\
\int \frac{d^dk\, k_m k_n}{(k^2+2kQ+M^2)^\alpha}&=&\frac{\pi^{d/2}}{
\Gamma(\alpha)(M^2-Q^2)^{\alpha-d/2}}\left[
Q_m Q_n\Gamma\left(\alpha-\frac d2\right)\right.
\nonumber\\&&
\left.+\frac 12 \delta_{mn}(M^2-Q^2)\Gamma\left(\alpha-1-\frac d2 \right)
\right]
\label{a3},\\
\int \frac{d^dk\, k_m k_n k_p}{(k^2+2kQ+M^2)^\alpha}&=&\frac{\pi^{d/2}}{
\Gamma(\alpha)(M^2-Q^2)^{\alpha-d/2}}\left[
-Q_m Q_n Q_p\Gamma\left(\alpha-\frac d2\right)\right.
\nonumber\\&&
-\frac 12 (\delta_{mn}Q_p+\delta_{np}Q_m+\delta_{pm}Q_n)
\nonumber\\&&\times\left.
(M^2-Q^2)\Gamma\left(\alpha-1-\frac d2 \right)
\right]
\label{a4},\\
\int \frac{d^dk\, k_m k_n k_p k_r}{(k^2+2kQ+M^2)^\alpha}&=&\frac{\pi^{d/2}}{
\Gamma(\alpha)(M^2-Q^2)^{\alpha-d/2}}\left[
Q_m Q_n Q_p Q_r\Gamma\left(\alpha-\frac d2\right)\right.
\nonumber\\&&
+\left.\frac 12 (\delta_{mn}Q_p Q_r+{\rm permutations})
(M^2-Q^2)\Gamma\left(\alpha-1-\frac d2 \right)\right.
\nonumber\\&&
+\frac14(\delta_{mn}\delta_{pr}+{\rm permutations})
\nonumber\\&&\times\left.
(M^2-Q^2)^2\Gamma\left(\alpha-2-\frac d2 \right)
\right].
\label{a5}
\end{eqnarray}

Each of the expressions (\ref{s1.26_}) has the corresponding representation in
the momentum space:
\begin{eqnarray}
\left[\frac{\partial_m}\square\delta^4(x-x')\right]^2
&\longrightarrow&
\int d^4k\frac{k_m(p+k)_n}{k^2(p+k)^2}\,,
\label{a6}\\
\left[\frac{\partial_m}\square\delta^4(x-x')\right]^3
&\longrightarrow&
\dint d^4k\frac{k_m(k-p_2)_n(k+p_1)_p}{k^2(k-p_2)^2(k+p_1)^2}\,,
\label{a7}\\
\left[\frac{\partial_m}\square\delta^4(x-x')\right]^4
&\longrightarrow&
\int d^4k\frac{k_m(k-p_2)_n(p_1+p_4+k)_p(p_1+k)_r}{k^2(k-p_2)^2
(k+p_1+p_4)^2(p_1+k)^2}\,.
\label{a8}
\end{eqnarray}
Using eqs. (\ref{a2}), (\ref{a3}), one can calculate the
integral (\ref{a6})
\be
\int d^4k\frac{k_m(p+k)_n}{k^2(p+k)^2}=-\pi^2(p^2)^{-\varepsilon}
\frac{\Gamma(\varepsilon)\Gamma^2(2-\varepsilon)}{\Gamma(4-2\varepsilon)}
\left[
p_mp_n+p^2\frac{\delta_{mn}}{2(1-\varepsilon)}
\right].
\label{a9}
\ee
In our calculations we are interested only in the divergent part
of the effective action. Therefore we should consider only the
divergent part of the expression (\ref{a6}) in the limit
$\varepsilon\to0$
\be
\left[\int d^4k\frac{k_m(p+k)_n}{k^2(p+k)^2} \right]_{div}
=-\frac{\pi^2}{6\varepsilon}[p_mp_n+p^2\frac{\delta_{mn}}2]\,.
\label{a10}
\ee
In particular,
\be
\left[\int d^4k\frac{k_n(p+k)_n}{k^2(p+k)^2} \right]_{div}
=-\frac{\pi^2}{2\varepsilon}p^2.
\label{a11}
\ee
The pole factor $1/\varepsilon$ appears here from the asymptotics of the gamma-function
$
\Gamma(\varepsilon)|_{\varepsilon\to0}=\frac1\varepsilon
-\gamma+O(\varepsilon)\,,
$ where $\gamma$ is Euler constant.

Similarly, using eqs. (\ref{a4}), (\ref{a5}),
we find the divergent parts of the remaining integrals
(\ref{a7}), (\ref{a8}):
\begin{eqnarray}
\left[
\int d^4k\frac{k_m(k-p_2)_n(k+p_1)_p}{k^2(k-p_2)^2(k+p_1)^2} \right]_{div}
&=&\frac{\pi^2}{12\varepsilon}
[\delta_{mn}(2p_1+p_2)_p-\delta_{pm}(p_1+2p_2)_n\nonumber\\&&
-\delta_{np}(p_1-p_2)_m],
\label{a12}\\
\left[
\int d^4k\frac{k_m(k-p_2)_n(p_1+p_4+k)_p(p_1+k)_r}{k^2(k-p_2)^2
(k+p_1+p_4)^2(p_1+k)^2}
\right]_{div}&=&\frac{\pi^2}{24\varepsilon}
(\delta_{mn}\delta_{pr}+\delta_{mp}\delta_{nr}+\delta_{mr}\delta_{np}).
\label{a13}
\end{eqnarray}

\subsection{Feynman graphs in SYM model}
The action (\ref{s1.22}) defines the Feynman  rules in the deformed gauge
model. The propagators have the standard form in the quantum field theory
listed in the following table\\
\vspace{0.1cm}

\begin{tabular}{|l|l|}
\hline\hline
Propagator & Line \\
\hline
$-\frac12\int d^4x \phi\square\bar\phi\rightarrow
\langle\phi(x)\bar\phi(x')\rangle=-\frac2\square\delta^4(x-x')$&
 \begin{picture}(81,19) (-1,-12)
    \SetWidth{1.0}
    \SetColor{Black}
    \Line(0,-11)(60,-11)
    \SetWidth{0.5}
    \Vertex(0,-11){1.41}
    \Vertex(60,-11){1.41}
    \Text(7,-9)[lb]{\small{\Black{$\phi$}}}
    \Text(50,-9)[lb]{\small{\Black{$\bar\phi$}}}
  \end{picture}\\ \hline
$i\int d^4x \Psi^{i\alpha}\partial_{\alpha\dot\alpha}\bar\Psi_i^{\dot\alpha}
\rightarrow
\langle\Psi^{i\alpha}(x)\bar\Psi_j^{\dot\alpha}(x')\rangle=
-\frac{2i\partial^{\alpha\dot\alpha}}{\square}\delta^4(x-x')\delta^i_j$&
  \begin{picture}(83,22) (-1,-9)
    \SetWidth{1.0}
    \SetColor{Black}
    \DashLine(0,-8)(60,-8){8}
    \SetWidth{0.5}
    \Vertex(60,-8){1.41}
    \Vertex(0,-8){1.41}
    \Text(6,-3)[lb]{\small{\Black{$\Psi$}}}
    \Text(52,-4)[lb]{\small{\Black{$\bar\Psi$}}}
  \end{picture}
\\ \hline
$-\frac12\int d^4x A_n\square A_n\rightarrow
\langle A_m(x)A_n(x')\rangle=-\frac2\square\delta^4(x-x')\delta_{mn}$&
  \begin{picture}(91,30) (-1,-8)
    \SetWidth{1.0}
    \SetColor{Black}
    \Photon(0,0)(70,0){4}{4}
    \Text(2,6)[lb]{\small{\Black{$A_m$}}}
    \Text(60,6)[lb]{\small{\Black{$A_n$}}}
    \SetWidth{0.5}
    \Vertex(0,0){1.41}
    \Vertex(70,0){1.41}
  \end{picture}
  \\
\hline\hline
\end{tabular}
\vspace{0.2cm}

The vertices defined by the action (\ref{s1.22}) have quite complicated
form. Schematically, they can be depicted as\\
\vspace{0.2cm}

\begin{tabular}{|l|l|}
\hline\hline
Interaction & Vertex\\\hline
$ -\frac12\dint d^4x
\dfrac{16I^3\square\bar\phi\partial_m\bar\phi
\partial_m\bar\phi}{1+4I\bar\phi}$ &
  \begin{picture}(100,86) (0,6)
    \SetWidth{0.5}
    \SetColor{Black}
    \Line(20,44)(25,49)
    \Line(25,44)(30,49)
    \Line(38,65)(43,70)
    \Line(38,20)(43,25)
    \Vertex(40,46){1.41}
    \SetWidth{1.0}
    \Line(40,46)(40,86)
    \Line(40,46)(0,46)
    \Line(40,46)(40,6)
    \SetWidth{2.0}
    \Line(40,46)(80,46)
    \Text(0,49)[lb]{\small{\Black{$\square\bar\phi$}}}
    \Text(70,49)[lb]{\small{\Black{$\frac1{1+4I\bar\phi}$}}}
    \Text(42,7)[lb]{\small{\Black{$\partial_m\bar\phi$}}}
    \Text(45,76)[lb]{\small{\Black{$\partial_m\bar\phi$}}}
  \end{picture}
\label{vertex1}\\\hline
$i\dint d^4x\dfrac{4IA_m\sigma_m{}^{\alpha}{}_{\dot\alpha}\bar\Psi^{i\dot\alpha}}{
1+4I\bar\phi}
(\sigma_n)_{\alpha\dot\beta}\partial_n
\left(\dfrac{\bar\Psi_i^{\dot\beta}}{1+4I\bar\phi} \right)$
&
  \begin{picture}(108,86) (0,6)
    \SetWidth{0.5}
    \Line(37,52)(46,48)
    \SetWidth{2.0}
    \SetColor{Black}
    \Line(40,46)(80,6)
    \Line(40,46)(80,86)
    \SetWidth{0.5}
    \Vertex(40,46){1.41}
    \SetWidth{1.0}
    \Photon(40,46)(0,46){3.5}{4}
    \Text(0,52)[lb]{\small{\Black{$A_m$}}}
    \DashLine(40,46)(40,86){8}
    \Text(78,74)[lb]{\small{\Black{$\frac{1}{1+4I\bar\phi}$}}}
    \Text(78,12)[lb]{\small{\Black{$\frac1{1+4I\bar\phi}$}}}
    \Text(42,9)[lb]{\small{\Black{$\bar\Psi$}}}
    \Text(42,76)[lb]{\small{\Black{$\bar\Psi$}}}
    \DashLine(40,46)(40,6){8}
  \end{picture}
  \label{vertex2}\\\hline
$
\begin{array}c
 \dint d^4x\left[A_n G_m\partial_n A_m- A_n G_n\partial_m
A_m-\varepsilon_{mnrs}G_mA_n\partial_r A_s\right]\\
=\dint d^4x X_{srn}(\bar\phi) A_n\partial_r A_s,\\
X_{srn}(\bar\phi)=G_s\delta_{rn}-G_n\delta_{rs}
-\varepsilon_{mnrs}G_m
\end{array}
$
&
 \begin{picture}(112,93)(0,11)
    \SetWidth{2.0}
    \SetColor{Black}
    \Line(0,53)(40,53)
    \SetWidth{0.5}
    \Line(55,72)(64,70)
    \Vertex(40,53){1.41}
    \SetWidth{1.0}
    \Photon(40,53)(80,93){3.5}{4}
    \Photon(40,53)(80,13){3.5}{4}
    \Text(82,88)[lb]{\small{\Black{$\partial_r A_s$}}}
    \Text(80,13)[lb]{\small{\Black{$A_n$}}}
    \Text(0,55)[lb]{\small{\Black{$X_{srn}(\bar\phi)$}}}
  \end{picture}
\label{vertex3}\\\hline
$\begin{array}c
i\dint d^4x\bar\Psi_i^{\dot\beta}(\sigma_m)_{\alpha\dot\alpha}
\partial_m\Psi^{i\alpha}\sum\limits_{n=2}^\infty(-4I\bar\phi)^n\\
=i\dint d^4x\bar\Psi_i^{\dot\beta}(\sigma_m)_{\alpha\dot\alpha}
\partial_m\Psi^{i\alpha}Y(\bar\phi),\\
Y(\bar\phi)=\sum\limits_{n=2}^\infty(-4I\bar\phi)^n
\end{array}$
&
  \begin{picture}(100,91) (0,11)
    \SetWidth{1.0}
    \SetColor{Black}
    \DashLine(0,91)(40,51){8}
    \DashLine(40,51)(0,11){8}
    \SetWidth{2.0}
    \Line(40,51)(80,51)
    \SetWidth{0.5}
    \Line(22,65)(30,65)
    \Vertex(40,51){1.41}
    \Text(10,11)[lb]{\small{\Black{$\bar\Psi$}}}
    \Text(70,54)[lb]{\small{\Black{$Y(\bar\phi)$}}}
    \Text(10,86)[lb]{\small{\Black{$\partial\Psi$}}}
  \end{picture}\\\hline\hline
\end{tabular}
\vspace{0.2cm}

Analyzing the propagators and vertices given above, one can observe that
there are only two types of nontrivial loop diagrams shown at Fig. 1.
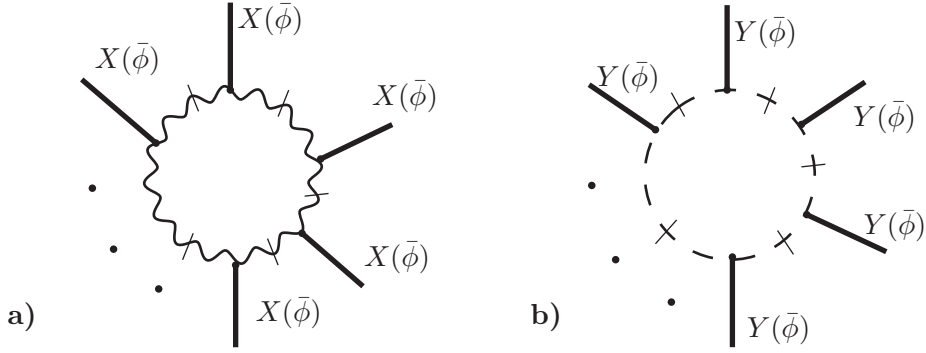
\begin{figure}[th]
\begin{center}
\begin{picture}(140,134) (28,-16)
    \SetWidth{0.5}
    \Line(72,73)(68,83)
    \Line(102,71)(106,81)
    \Line(112,41)(121,42)
    \Line(98,25)(102,15)
    \Line(70,27)(66,17)
    \SetWidth{1.0}
    \SetColor{Black}
    \PhotonArc(85,49)(31.11,-45,315){2.5}{16}
    \Text(33,87)[lb]{\small{\Black{$X(\bar\phi)$}}}
    \SetWidth{2.0}
    \Line(86,15)(86,-16)
    \Line(111,27)(134,7)
    \Line(118,55)(145,68)
    \Line(56,61)(28,85)
    \Line(84,81)(84,114)
    \SetWidth{0.5}
    \Vertex(32,44){1.41}
    \Text(135,12)[lb]{\small{\Black{$X(\bar\phi)$}}}
    \SetWidth{0.5}
    \Vertex(56,61){1.41}
    \Vertex(84,81){1.41}
    \Vertex(118,55){1.41}
    \Vertex(111,27){1.41}
    \Vertex(86,15){1.41}
    \Text(87,102)[lb]{\small{\Black{$X(\bar\phi)$}}}
    \Text(138,73)[lb]{\small{\Black{$X(\bar\phi)$}}}
    \Text(94,-9)[lb]{\small{\Black{$X(\bar\phi)$}}}
    \Vertex(40,21){1.41}
    \Vertex(57,6){1.41}
    \Text(0,-10)[lb]{\small{\Black{\bf a)}}}
  \end{picture}
\qquad\qquad
   \begin{picture}(134,129) (31,-21)
    \SetWidth{0.5}
    \Line(68,66)(63,75)
    \Line(98,67)(102,77)
    \Line(112,47)(121,48)
    \Line(103,23)(109,14)
    \Line(65,27)(57,18)
    \SetWidth{1.0}
    \SetColor{Black}
    \DashCArc(85,44)(32.02,129,489){8}
    \SetWidth{2.0}
    \Line(57,61)(32,78)
    \Line(84,76)(84,108)
    \Line(112,63)(136,79)
    \Line(114,29)(144,15)
    \Text(35,75)[lb]{\small{\Black{$Y(\bar\phi)$}}}
    \Text(87,92)[lb]{\small{\Black{$Y(\bar\phi)$}}}
    \Text(132,60)[lb]{\small{\Black{$Y(\bar\phi)$}}}
    \Line(86,13)(86,-21)
    \SetWidth{0.5}
    \Vertex(57,61){1.41}
    \Vertex(84,76){1.41}
    \Vertex(112,63){1.41}
    \Vertex(114,29){1.41}
    \Vertex(86,13){1.41}
    \Vertex(33,40){1.41}
    \Vertex(42,12){1.41}
    \Vertex(63,-4){1.41}
    \Text(92,-20)[lb]{\small{\Black{$Y(\bar\phi)$}}}
    \Text(136,19)[lb]{\small{\Black{$Y(\bar\phi)$}}}
    \Text(10,-15)[lb]{\small{\Black{\bf b)}}}
  \end{picture}
\end{center}
\caption{Two types of one-loop diagrams with vector and spinor
internal lines.} \label{diag2}
\end{figure}
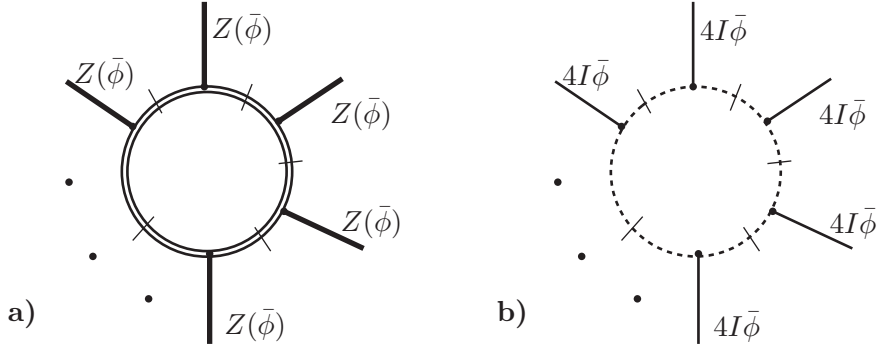
\begin{figure}[ht]
\begin{center}
   \begin{picture}(134,129) (31,-21)
    \SetWidth{0.5}
    \Line(68,66)(63,75)
    \Line(98,67)(102,77)
    \Line(112,47)(121,48)
    \Line(103,23)(109,14)
    \Line(65,27)(57,18)
    \SetWidth{1.0}
    \SetColor{Black}
    \CArc(85,44)(32.1,129,489)
    \CArc(85,44)(30,129,489)
    \SetWidth{2.0}
    \Line(57,61)(32,78)
    \Line(84,76)(84,108)
    \Line(112,63)(136,79)
    \Line(114,29)(144,15)
    \Text(35,75)[lb]{\small{\Black{$Z(\bar\phi)$}}}
    \Text(87,92)[lb]{\small{\Black{$Z(\bar\phi)$}}}
    \Text(132,60)[lb]{\small{\Black{$Z(\bar\phi)$}}}
    \Line(86,13)(86,-21)
    \SetWidth{0.5}
    \Vertex(57,61){1.41}
    \Vertex(84,76){1.41}
    \Vertex(112,63){1.41}
    \Vertex(114,29){1.41}
    \Vertex(86,13){1.41}
    \Vertex(33,40){1.41}
    \Vertex(42,12){1.41}
    \Vertex(63,-4){1.41}
    \Text(92,-20)[lb]{\small{\Black{$Z(\bar\phi)$}}}
    \Text(136,19)[lb]{\small{\Black{$Z(\bar\phi)$}}}
    \Text(10,-15)[lb]{\small{\Black{\bf a)}}}
  \end{picture}
\qquad\qquad
   \begin{picture}(134,129) (31,-21)
    \SetWidth{0.5}
    \Line(68,66)(63,75)
    \Line(98,67)(102,77)
    \Line(112,47)(121,48)
    \Line(103,23)(109,14)
    \Line(65,27)(57,18)
    \SetWidth{1.0}
    \SetColor{Black}
    \DashCArc(85,44)(32.02,129,489){2}
    \SetWidth{1.0}
    \Line(57,61)(32,78)
    \Line(84,76)(84,108)
    \Line(112,63)(136,79)
    \Line(114,29)(144,15)
    \Text(35,75)[lb]{\small{\Black{$4I\bar\phi$}}}
    \Text(87,92)[lb]{\small{\Black{$4I\bar\phi$}}}
    \Text(132,60)[lb]{\small{\Black{$4I\bar\phi$}}}
    \Line(86,13)(86,-21)
    \SetWidth{0.5}
    \Vertex(57,61){1.41}
    \Vertex(84,76){1.41}
    \Vertex(112,63){1.41}
    \Vertex(114,29){1.41}
    \Vertex(86,13){1.41}
    \Vertex(33,40){1.41}
    \Vertex(42,12){1.41}
    \Vertex(63,-4){1.41}
    \Text(92,-20)[lb]{\small{\Black{$4I\bar\phi$}}}
    \Text(136,19)[lb]{\small{\Black{$4I\bar\phi$}}}
    \Text(10,-15)[lb]{\small{\Black{\bf b)}}}
  \end{picture}
\end{center}
\caption{Two types of one-loop diagrams with scalar and spinor
internal lines.} \label{diag3}
\end{figure}
Both these diagrams have arbitrary numbers of external lines. The effective
action corresponding to the diagram a) is calculated in
Sect. 2. The sum of diagrams b) makes the trivial
contribution to the effective action. Indeed, it corresponds to the
following one-loop effective action
\begin{eqnarray}
\Gamma_\Psi&=&\Tr\ln\frac{\delta^2S_\Psi}{\delta\Psi_i^\alpha(x)
\delta\bar\Psi^{j\dot\alpha}}
=\Tr\ln\left[-i\frac{\delta^i_j(\sigma_n)_{\alpha\dot\alpha}
\partial_n\delta^4(x-x')}{1+4I\bar\phi}
 \right]\nonumber\\
&=&2\Tr\ln[(\sigma_n)_{\alpha\dot\alpha}
\partial_n\delta^4(x-x')]+2\Tr\ln\left[\frac{\delta^4(x-x')}{
1+4I\bar\phi} \right]\simeq0\,.
\label{b1}
\end{eqnarray}
Both terms in the second line of eq. (\ref{b1}) are trivial.

\subsection{Feynman graphs in hypermultiplet model}

Feynman rules in the deformed hypermultiplet model are defined by the action
(\ref{s2.4}). The propagators have the standard form
\vspace{0.2cm}

\begin{tabular}{|l|l|}
\hline\hline
Propagator & Line \\\hline
$-\frac12 \int d^4x f^{ak}\square f_{ak}
\rightarrow\langle f^{ak}(x)f_{a'k'}(x')\rangle
=-\frac2{\square}\delta^4(x-x')\delta^a_{a'}\delta^k_{k'}$
&
 \begin{picture}(81,19) (-1,-12)
    \SetWidth{1.0}
    \SetColor{Black}
    \Line(0,-10)(60,-10)
    \Line(0,-12)(60,-12)
    \Vertex(0,-11){1.41}
    \Vertex(60,-11){1.41}
    \Text(0,-9)[lb]{\small{\Black{$f^{ak}$}}}
    \Text(40,-9)[lb]{\small{\Black{$f_{a'k'}$}}}
  \end{picture}\\ \hline
$\frac i2\int d^4x \rho^{\alpha a}\partial_{\alpha\dot\alpha}
\chi_a^{\dot\alpha}\rightarrow
\langle\rho^{\alpha a}(x)\chi^{\dot\alpha}_b(x')
\rangle=-i\frac{\partial^{\alpha\dot\alpha}}\square
\delta^4(x-x')\delta^a_b$
&
 \begin{picture}(81,19) (-1,-12)
    \SetWidth{1.0}
    \SetColor{Black}
    \DashLine(0,-11)(60,-11){2}
    \Vertex(0,-11){1.41}
    \Vertex(60,-11){1.41}
    \Text(0,-9)[lb]{\small{\Black{$\rho^{\alpha a}$}}}
    \Text(40,-9)[lb]{\small{\Black{$\chi^{\dot\alpha}_b$}}}
  \end{picture}\\ \hline\hline
\end{tabular}
\vspace{0.2cm}

The vertices defined by the action (\ref{s2.4}) are given in the following table
\\
\begin{tabular}{|l|l|}
\hline\hline
Interaction & Vertex \\ \hline
$\begin{array}c
\frac12\dint d^4x(8I\bar\phi+16I^2\bar\phi^2)
\partial_m f^{ak}\partial_m f_{ak}\\
=\frac12\dint d^4x Z(\bar\phi)
\partial_m f^{ak}\partial_m f_{ak},\\
Z(\bar\phi)=8I\bar\phi+16I^2\bar\phi^2
\end{array}
$
&
\begin{picture}(135,55) (38,18)
    \SetWidth{2.0}
    \SetColor{Black}
    \Line(40,44)(90,44)
    \SetWidth{1.0}
    \Line(89,46)(114,71)\Line(91,43)(117,68)%%JaxoDrawID
    \Line(91,45)(116,20)\Line(89,42)(114,17)%%JaxoDrawID:
    \Vertex(90,44){3}
    \SetWidth{0.5}
    \Line(93,55)(108,55)
    \Line(92,35)(107,35)
    \Text(40,50)[lb]{\small{\Black{$Z(\bar\phi)$}}}
    \Text(115,57)[lb]{\small{\Black{$\partial_m f^{ak}$}}}
    \Text(115,23)[lb]{\small{\Black{$\partial_m f_{ak}$}}}
\end{picture}
\\ \hline
$\frac i2\dint d^4x 4I\bar\phi\rho^{\alpha a}
\partial_{\alpha\dot\alpha}\chi_a^{\dot\alpha}$&
\begin{picture}(95,55) (38,18)
    \SetWidth{1.0}
    \SetColor{Black}
    \Line(40,44)(90,44)
    \SetWidth{1.0}
    \DashLine(90,44)(115,70){2}
    \DashLine(90,44)(115,20){2}
    \Vertex(90,44){2}
    \SetWidth{0.5}
    \Line(95,35)(105,35)
    \Text(42,50)[lb]{\small{\Black{$4I\bar\phi$}}}
    \Text(110,55)[lb]{\small{\Black{$\rho^\alpha_a$}}}
    \Text(110,25)[lb]{\small{\Black{$\partial_{\alpha\dot\alpha}
      \chi_a^{\dot\alpha}$}}}
\end{picture}
\\ \hline
$4i\dint d^4x \bar\Psi^{\dot\alpha}_k\rho^\alpha_a\partial_{\alpha\dot\alpha}
f^{ak}$&
\begin{picture}(105,55) (38,18)
    \SetWidth{1.0}
    \SetColor{Black}
    \DashLine(40,44)(90,44){8}
    \SetWidth{1.0}
    \DashLine(90,44)(115,70){2}
    \Line(91,45)(116,21)\Line(89,43)(114,19)
    \Vertex(90,44){2}
    \SetWidth{0.5}
    \Line(95,35)(105,35)
    \Text(42,50)[lb]{\small{\Black{$\bar\Psi_k^{\dot\alpha}$}}}
    \Text(110,55)[lb]{\small{\Black{$\rho_a^\alpha$}}}
    \Text(110,25)[lb]{\small{\Black{$\partial_{\alpha\dot\alpha} f^{ak}$}}}
\end{picture}
\\ \hline
$2i\dint d^4x I\rho^{\alpha a} A_m \partial_m\rho_{\alpha a}$&
\begin{picture}(105,55) (38,18)
    \SetWidth{1.0}
    \SetColor{Black}
    \Photon(40,44)(90,44){3}{4}
    \SetWidth{1.0}
    \DashLine(90,44)(115,70){2}
    \DashLine(90,44)(115,20){2}
    \Vertex(90,44){2}
    \SetWidth{0.5}
    \Line(95,35)(105,35)
    \Text(42,50)[lb]{\small{\Black{$A_m$}}}
    \Text(110,55)[lb]{\small{\Black{$\rho^{\alpha a}$}}}
    \Text(110,25)[lb]{\small{\Black{$\partial_m\rho_{\alpha a}$}}}
\end{picture}
\\ \hline
$i\dint d^4x I\rho^{\beta a}\rho_a^\alpha\partial_{(\alpha\dot\alpha}
A^{\dot\alpha}_{\beta)}$&
\begin{picture}(105,55) (38,18)
    \SetWidth{1.0}
    \SetColor{Black}
    \Photon(40,44)(90,44){3}{4}
    \SetWidth{1.0}
    \DashLine(90,44)(115,70){2}
    \DashLine(90,44)(115,20){2}
    \Vertex(90,44){2}
    \SetWidth{0.5}
    \Line(65,40)(75,50)
    \Text(42,50)[lb]{\small{\Black{$\partial_{(\alpha\dot\alpha}A_{
    \beta)}^{\dot\alpha}$}}}
    \Text(110,55)[lb]{\small{\Black{$\rho^{\beta a}$}}}
    \Text(110,25)[lb]{\small{\Black{$\rho^\alpha_a$}}}
\end{picture}
\\ \hline\hline
\end{tabular}
\vspace{0.2cm}

Like in the gauge model, one can observe that there
are only two types of nontrivial diagrams shown in Fig. 2. The computation of these
diagrams is considered in Sect. 3. Note that the diagram with the fermionic loop
give only trivial contribution to the effective action, see eq. (\ref{s2.5}).

\end{document}